\documentclass[aps,prb,showpacs,twocolumn]{revtex4}
\usepackage{amsmath,amssymb,mathrsfs,bm,feynmf,setspace ,txfonts}
\usepackage{graphicx}
\usepackage{color}
\usepackage{hyperref}
\newcommand{\al}{\alpha}
\newcommand{\be}{\beta}

\newcommand{\de}{\delta}
\newcommand{\e}{\epsilon}

\newcommand{\thi}{\theta}

\newcommand{\ka}{\kappa}
\newcommand{\la}{\lambda}
\newcommand{\La}{\Lambda}
\newcommand{\mi}{\mu}

\newcommand{\p}{\pi}

\newcommand{\s}{\sigma}

\newcommand{\f}{\phi}

\newcommand{\w}{\omega}
\newcommand{\W}{\Omega}
\newcommand{\De}{\Delta}
\newcommand{\G}{\Gamma}
\renewcommand{\S}{\Sigma}

\newcommand{\pd}{\partial}

\newcommand{\round}[1]{\left( #1 \right)}
\renewcommand{\square}[1]{\left[ #1 \right]}
\newcommand{\abs}[1]{\left| #1 \right|}
\newcommand{\cvec}[2]{\round{\begin{array}{c} #1 \\ #2 \end{array}}}

\newcommand{\mat}[4]{\left(\begin{array}{cc}#1&#2\\#3&#4\end{array}\right)}

\newcommand{\rvec}[2]{\round{\begin{array}{cc}#1&#2\end{array}}}

\newcommand{\ang}[1]{\left\langle #1 \right\rangle}

\newcommand{\beq}{\begin{equation}}
\newcommand{\eeq}{\end{equation}}
\newcommand{\Beq}{\begin{eqnarray}}
\newcommand{\Eeq}{\end{eqnarray}}
\newcommand{\bml}{\begin{multline}}

\newcommand{\eeqm}{\end{multline}}

\newcommand{\bsp}{\begin{split}}
\newcommand{\esp}{\end{split}}

\renewcommand{\b}[1]{{\bf #1}}
\renewcommand{\t}{\tilde}
\newcommand{\Te}{T_{\rm eff}}

\newcommand{\cL}{\mathcal L}
\renewcommand{\t}{\tilde}
\newcommand{\req}[1]{Eq.~(\ref{#1})}

\newcommand{\nn}{\nonumber\\}

\DeclareMathOperator{\Tr}{Tr}

\DeclareMathOperator{\sgn}{sgn}

\begin{document}

\title{Nonequilibrium quantum criticality in bilayer itinerant ferromagnets}
\author{So Takei$^1$, William Witczak-Krempa$^2$ and Yong Baek Kim$^{2,3}$}
\affiliation{
$^1$Max-Planck-Institut f\"{u}r Festk\"{o}rperforschung, D-70569 Stuttgart, Germany\\
$^2$Department of Physics, University of Toronto, Toronto, Ontario M5S 1A7, Canada\\
$^3$School of Physics, Korea Institute for Advanced Study, Seoul 130-722, Korea}
\date{\today}
\pacs{03.65.Yz, 05.30.-d, 71.10.-w}

\begin{abstract}
We present a theory of nonequilibrium quantum criticality in a coupled bilayer system of itinerant 
electron magnets. The model studied consists of the first layer subjected to an inplane current 
and open to an external substrate. The second layer is subject to no direct external drive, but 
couples to the first layer via short-ranged spin exchange interaction. No particle exchange is 
assumed between the layers. Starting from a microscopic fermionic model, we derive an effective action
in terms of two coupled bosonic fields which are related to the magnetization fluctuations of
the two layers. When there is no interlayer coupling, the two bosonic modes possess different 
dynamical critical exponents $z$ with $z=2$ ($z=3$) for the first (second) layer. 
This results in multi-scale quantum criticality in the coupled system.
It is shown that the linear coupling between the two fields leads to a low-energy 
fixed point characterized by the larger dynamical critical exponent $z=3$.
We compute the correlation length in the quantum disordered and quantum critical regimes
for both the nonequilibrium case and in thermal equilibrium where the whole system is held at a common
temperature $T$. We identify an effective temperature scale $T_{\rm eff}$ with which we define
a quantum-to-classical crossover that is in exact analogy with the thermal equilibrium case but
with $T$ replaced by $T_{\rm eff}$. However, we find that the leading correction to the correlation length
in the quantum critical regime scales differently with respect to $T$ and $T_{\rm eff}$. We also note that
the current in the lower layer generates a drift of the magnetization fluctuations, manifesting itself as a parity-breaking
contribution to the effective action of the bosonic modes. In this sense,
the nonequilibrium drive in this system plays a role which is distinct from $T$ in the thermal
equilibrium case.
We also derive the stochastic dynamics obeyed by the critical fluctuations in the quantum critical 
regime and find that they do not fall into the previously identified dynamical universality classes. 
\end{abstract}

\maketitle

\section{Introduction}
Understanding the role of nonequilibrium drives on systems tuned near a quantum critical point has 
received growing attention in recent years. One of the issues is to understand to what extent the 
concept of universality persists when systems depart far from equilibrium. Several works 
considered universal scaling behaviour in nonlinear transport properties close to a superconductor-insulator 
quantum critical point\cite{noise,cond1,cond2}. Universal scaling of current in response to an external field 
was also studied for an itinerant electron system near a magnetic quantum critical point\cite{curr}.
More recently, universal scaling functions of the non-linear conductance and fluctuation dissipation
ratios were obtained for a magnetic single-electron transistor near quantum criticality\cite{ks1,ks2}.

Complimentary works have studied the effects of current flow on 2D itinerant electron systems near a 
ferromagnetic-paramagnetic quantum critical point by generalizing the perturbative renormalization group 
to nonequilibrium situations\cite{aditi1,aditi2}. These works developed scaling theories which showed 
that the leading effect of the nonequilibrium probe is to act as an effective temperature, $T_{\rm eff}$. 
As in the corresponding equilibrium theory\cite{hertz,millis}, $T_{\rm eff}$ was found to be a relevant 
perturbation at the equilibrium gaussian fixed point and to be responsible for inducing a quantum-to-classical 
crossover similar to the one triggered by thermal fluctuations.  A large-$T_{\rm eff}$ quantum critical regime, 
analogous to the corresponding regime in equilibrium quantum criticality, was also identified. In this regime, 
the long-time, long-wavelength behaviour of the order parameter fluctuations was found to obey Langevin dynamics
similar to one of the dynamical universality classes considered by Hohenberg and Halperin\cite{handh}.
This has led to an important identification of a nonequilibrium dynamical universality class.

The form of the long-time, long-wavelength dynamics obeyed by order parameter fluctuations depends intimately 
on the geometry of the system and on the manner in which the nonequilibrium perturbation is applied. For a 2D 
system, the orientation of the current flow with respect to the plane of the system influences the dynamics. 
When inversion symmetry is broken\cite{aditi2}, for instance with the application of current parallel to the 
plane, the dynamics obeyed by the order parameter fluctuations differs from the case where 
the current is applied normal to the plane. Introducing particle exchange by tunnel-coupling the system 
to an external reservoir also affects the dynamical critical phenomena. A system, initially characterized by 
conserved order parameter fluctuations in the closed case, must be described by non-conserved fluctuations once 
it is opened. For itinerant ferromagnets\cite{millis}, this corresponds to a change in the dynamical critical
exponent from $z=3$ to $z=2$.\cite{aditi1} The sensitivity of order parameter dynamics to system geometry and 
the form of nonequilibrium perturbation opens up the possibility to explore a variety of nonequilibrium 
dynamical universality classes and contribute to the general understanding of nonequilibrium quantum cricality.

In this work, we consider nonequilibrium quantum criticality in a coupled bilayer system of itinerant ferromagnets 
tuned close to a spin-density wave instability. The two itinerant ferromagnets are separated by a 3D insulating 
barrier (see Fig.\ref{fig:sys}), and are coupled via short-ranged spin exchange interaction. We consider the case
where there is no particle exchange between the layers. Departure from equilibrium is achieved by driving one of 
the layers with an in-plane electrical current. The resulting drift\cite{aditi2,aditi3} of the spin fluctuations, in turn, 
causes the fluctuations in the other layer to also drift, ultimately driving the latter out of equilibrium. A steady state is
established by tunnel-coupling the driven layer to an external bath. 

We show that the system can be characterized by two bosonic fields, which simply relate to the physical spin 
fluctuations of the two layers, and are coupled at the gaussian level via the interlayer exchange 
interaction. The crucial complexity of the problem comes from the fact that
the two fields possess different dynamics with different dynamical exponents, $z$,
in the absence of interlayer coupling.
This is a direct consequence of the fact that while one layer is open to an external 
substrate ($z=2$ mode), the other remains closed from any external bath ($z=3$ mode). As such, the bilayer system 
exhibits multi-scale quantum criticality. We find that, because the fields couple linearly, the infrared properties 
of the system are governed by a $z=3$ fixed point.  

We present a perturbative renormalization group analysis of the system both in and out of equilibrium. 
In the equilibrium case the full system is held at a common temperature, $T$. Out of equilibrium, we 
introduce an effective temperature, $T_{\rm eff}$ (as in Ref. \onlinecite{aditi2}), which parametrizes the 
decoherence induced by the nonequilibrium drive.
We study the flow of the system in the vicinity of the $z=3$ equilibrium gaussian fixed point. We identify the 
line of crossover from the quantum disordered to the quantum critical regime in terms of the bare parameters 
of the system, and compute both temperature and nonequilibrium corrections to the correlation length of the 
critical fluctuations in the quantum critical regime. 
We find that while the temperature corrections to the correlation length in thermal equilibrium gain two contributions that reflect the presence of the $z=2$ and $z=3$ dynamics, the nonequilibrium contribution 
contains only one correction 
that reflects $z=2$ physics. This result signifies the fact that the nonequilibrium drive is applied only 
to the bottom layer whose corresponding fluctuations in the decoupled limit possess $z=2$ dynamics. 
This shows that the energy scale identified as effective temperature in a previous work 
(Ref. \onlinecite{aditi2}) does not strictly apply to this problem.

We also consider the Langevin dynamics obeyed by the critical fluctuations in the quantum critical regime 
where the system effectively becomes classical and the quantum fluctuations can be integrated out. Here, the 
analysis is carried out in the eigenbasis that diagonalizes the gaussian effective
action. Because one of the eigenmodes remains massive at the critical point, the long-time dynamics of the full
system can be described by a single field. We solve the Langevin equation for this eigenmode both
in and out of equilibrium. While $z=2$ physics seem to play no role in the long-time dynamics of
the critical eigenmode in equilibrium, its effect is important in the nonequilibrium case. In the latter case, 
the Langevin dynamics display a hybrid effect of both $z=2$ and $z=3$ physics. In either case, the obtained
dynamics differ from any of the dynamical unversality classes considered in Ref. \onlinecite{handh}.

The paper is organized as follows. In Sec. \ref{system}, we begin by introducing the bilayer system, and present 
a theory to model the system. By integrating out all fermionic degrees of freedom, the effective two-field 
bosonic action  is derived in Sec. \ref{effaction}. We provide a brief mean-field analysis of the effective 
action, and establish the fixed point in Sec. \ref{precons} for the perturbative renormalization group analysis 
which follows in Sec. \ref{RGA}. The correlation length is calculated in Sec. \ref{CLs}, and the dynamics obeyed 
by the critical fluctuations in the quantum critical regime is discussed in Sec. \ref{langeq}. Finally, we 
conclude in Sec. \ref{conclude}.

\section{System and Model}\label{system}
\begin{figure}[t]
\begin{center}
\includegraphics[scale=0.8]{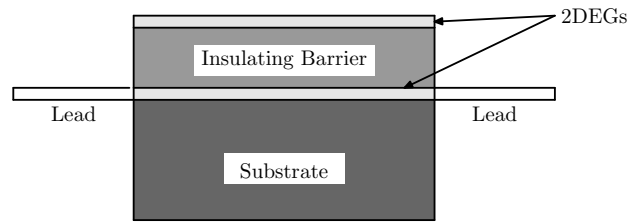}
\caption{\label{fig:sys} Edge-on view of the model system considered
in this work. The two itinerant electron films (2DEGs) are coupled via short-ranged spin exchange 
interaction. The bottom 2DEG is tunnel-coupled to a substrate and is driven out of equilibrium by a 
uniform electric field $\b E$ applied in the inplane direction. The top layer in turn is driven out of 
equilibrium by its interaction with the lower layer. There is no tunneling of electrons between the two 
2DEGs.}
\end{center}
\end{figure}

The system of interest, shown in Fig.\ref{fig:sys}, consists of two 2D itinerant electron systems 
which are tuned close to a spin-density wave instability\cite{hertz,millis}. 
Throughout this work, we assume the systems to be paramagnetic but are nearly unstable
toward an itinerant ferromagnetic state with an order parameter symmetry of Ising nature. 
We assume no particle exchange between the two layers, but allow magnetic fluctuations in the 
layers to interact via short-ranged ferromagnetic spin exchange. The bottom layer is tunnel-coupled 
to a substrate and is driven out of equilibrium by a uniform electric field $\b E$ applied 
in the inplane direction. 

The total Hamiltonian of the system is given by
\beq
H=H_b+H_t+H_{b-t}+H_{sub}+H_{b-sub},
\eeq
where $b$ ($t$) labels the bottom (top) layer and $sub$ denotes the substrate. The Hamiltonian for
each layer has a kinetic part and an onsite ferromagnetic exchange interaction term
with a common interaction strength $U$. They can be written as
\Beq
\label{hbb}
H_b&=&\sum_{\b{k},\s} \e_{\b{k}-e\b{A}/c}
c_{b{\b k},\s}^\dag c_{b{\b k},\s} - \frac{U}{2}\sum_i (S^z_{b,i})^2;\\
\label{htt}
H_t&=&\sum_{\b{k},\s} \e_{{\b k}}c_{t{\b k},\s}^\dag c_{t{\b k},\s} - \frac{U}{2}\sum_i (S^z_{t,i})^2,
\Eeq
where $c_{b(t){\b k},\s}$ is the annihilation operator for the bottom (top) fermions 
with inplane momentum ${\b k}$ and spin $\s$. The usual quadratic dispersion, 
$\e_{\b k}=|{\b k}|^2/2m$, is assumed in both layers. $S^z_{\al,i}=\sum_{\s}\sigma
c_{\al i\s}^\dag c_{\al i\s'}$ is the magnetization for layer $\al$, where $i$ labels 
the inplane position, and the usual factor of 1/2 was absorbed in the interaction $U$.
$e$ is the electron charge. We use a gauge in which the electromagnetic potential 
${\b A}=-c\b E t$, and set $\hbar=1$. 

Eqs.(\ref{hbb}) and (\ref{htt}) have been written in a form which explicitly breaks 
SU(2)-symmetry. Here, we are assuming that the itinerant magnets possess an easy 
axis so that
their magnetization prefers to point up or down with respect to a certain crystal axis. 
Indeed, strong spin-orbit coupling in some itinerant magnets is known give rise to 
pronounced magnetocrystalline anisotropy\cite{roschrev,stewart}. 
In this work, we also focus solely on the disordered side of the phase diagram where
number of components for each of the order parameters are not expected to influence the results in an 
essential way\cite{hertz,millis}.

The Hamiltonian describing the inter-layer interaction 
reads
\beq
H_{b-t}=-J\sum_i S^z_{b,i}S^z_{t,i}\ ,
\eeq
where $J>0$ is the strength of the ferromagnetic exchange interaction. The interaction may be 
mediated by magnetic fluctuations in the central paramagnetic insulator. We assume that these 
fluctuations possess long-range correlations in the out-of-plane direction (on the order of its thickness), 
but short-ranged correlations in the inplane direction, which justifies the locality of the interaction in the 
inplane direction. We refer the reader to Appendix \ref{spinXc} for further details on how the paramagnetic 
fluctuations of the central insulator can give rise to an effective spin exchange interaction between 
the two layers. The substrate Hamiltonian and the tunneling Hamiltonian, which describes the 
coupling between the bottom layer and the substrate, are given by
\Beq
H_{sub}&=&\sum_{\b k,k_z,\s} 
\e_{{\b k},k_z}a_{{\b k},k_z,\s}^\dag a_{{\b k},k_z,\s};\\
\label{hbsub}
H_{b-sub}&=& \zeta\sum_{\b k,k_z,\s}\,a^\dag_{\b k,k_z,\s}c_{b\b k,\s} + {\rm h.c.},
\Eeq
where $a_{{\b k},k_z,\s}$ is the annihilation operator for the subtrate fermions, 
and $\zeta$ parametrizes the tunneling strength between the layer and the substrate. The substrate
is modeled as a Fermi-liquid bath with $\e_{{\b k},k_z}=\e_{\b k}+\e(k_z)$,
where $k_z$ labels the out-of-plane momentum, and $\e(k_z)$ is some dispersion not necessarily
quadratic. The essential features of the bath are its density of states, $\rho$, which implies a broadening
in the electronic levels associated with the bottom layer, $\G=\p\rho\zeta^2$ (the
escape rate of electrons from the bottom layer to the substrate), and its 
resistivity, which we assume to be very high relative to that of the layer so we may
couple the electric field only to the bottom layer.

\section{Effective Keldysh bosonic action}
\label{effaction}
To harness the nonequilibrium nature of the problem, we formulate our model using the Keldysh path 
integral formalism\cite{kamenev}. We begin by writing the Keldysh partition function
\beq
\mathcal{Z}^K=\int\mathcal{D}[\bar{c}^\pm,c^\pm,\bar{a}^\pm,a^\pm]e^{i\mathcal{S}^+-i\mathcal{S}^-},
\eeq
where $+$ and $-$ label the action for the forward and backward parts of the time-loop
contour, respectively. The full action is given by 
\beq
\mathcal{S}^\ka=\mathcal{S}^\ka_b+\mathcal{S}^\ka_t+\mathcal{S}^\ka_{b-t}+\mathcal{S}^\ka_{sub}
+\mathcal{S}^\ka_{b-sub},
\eeq
where
\begin{multline}
\mathcal{S}^\ka_b=\int d^3x\, \bar{c}^\ka_{b\s}\square{i\pd_t-
\frac{\round{i\pd_{\b r}+e{\b A}/c}^2}{2m}+\mu_b}c^\ka_{b\s}\\
+\frac{U}{2}\int d^3x\,(S_{b}^{z,\ka})^2,
\end{multline}
\begin{multline}
\mathcal{S}^\ka_t=\int d^3x\, \bar{c}^\ka_{t\s}\square{i\pd_t+\frac{\pd_{\b r}^2}{2m}
+\mu_t}c^\ka_{t\s}
+\frac{U}{2}\int d^3x\,(S_{t}^{z,\ka})^2,
\end{multline}
\beq
\mathcal{S}^\ka_{b-t}=J\int d^3x\, S_{b}^{z,\ka}S_{t}^{z,\ka},
\eeq
\beq
\mathcal{S}^\ka_{sub}=\int d^3x dz\, \bar{a}^\ka_{\s}\square{i\pd_t+\frac{\pd_{\b r}^2}{2m}
-\e(-i\pd_z)+\mu_s}a^\ka_{\s},
\eeq
\beq
\mathcal{S}^\ka_{b-sub}=-\zeta\int d^3x\,\round{\bar{a}_{\s,z=0}^\ka c_{b\s}^\ka
+\bar{c}_{b\s}^\ka a_{\s,z=0}^\ka}.
\eeq
We are using $x=(t,\b r)$, $\b r$ being the inplane coordinate; $z$ denotes the out of plane direction.
All fermionic fields are now expressed in terms of Grassmann variables. We have omitted position 
and time dependences from all fields for brevity, and assumed summation over the spin. $\ka=\pm$ label a branch 
of the Keldysh contour on which the corresponding field resides. $\mu_b$, $\mu_t$ and $\mu_s$ are the 
chemical potentials of the bottom layer, top layer and substrate, respectively.

\subsection{Free fermionic action}
\label{freefermionaction}
The free fermionic action in Keldysh space can be obtained by setting $U=J=0$. After performing 
a standard change of basis\cite{kamenev},
\beq\bsp
c^1&=\frac{c^++c^-}{\sqrt{2}},\quad c^2=\frac{c^+-c^-}{\sqrt{2}},\\
\bar{c}^1&=\frac{\bar{c}^+-\bar{c}^-}{\sqrt{2}},\quad \bar{c}^2=\frac{\bar{c}^++\bar{c}^-}{\sqrt{2}},
\end{split}\eeq
the Keldysh action for layer $\al$ is simply given by
\begin{multline}
\label{free_action}
\mathcal{S}_{\al ,0} = \int_{k,\s}\rvec{\bar{c}^1_{\al  k\s}}{\bar{c}_{\al  k\s}^2}
\mat{[G^{-1}_{\al  k}]^R}{[G^{-1}_{\al  k}]^K}{0}{[G^{-1}_{\al  k}]^A}\cvec{c^1_{\al  k\s}}{c^2_{\al  k\s}}.
\end{multline}
Here, $k=(\w,{\b k})$ labels the energy-momentum three-vector and we have introduced the notation
$\int_{k,\s}=\sum_\s\int\frac{d^2{\b k}}{(2\p)^2}\frac{d\w}{2\p}$. 
The top layer electron Green's functions correspond to those of free equilibrium electrons,
\Beq
G^{R(A)}_{tk}&=&\frac{1}{\w-\xi^t_{\b k}\pm i0^+},\nn
G^K_{tk}&=&-2\p i\sgn(\w)\de(\w-\xi^t_{\b k}),
\Eeq
where $\xi^t_{\b k}=\e_{\b k}-\mi_t$. We note that the Keldysh Green's function is given in its 
zero-temperature form. 

The bottom layer is affected by both the electric field and the substrate, and this substantially
modifies the Green's functions. Derivation of the Green's functions will be briefly outlined here. More 
detailed derivations can be found in Ref. \onlinecite{aditi2} and Ref. \onlinecite{aditi3}. The first step is to 
integrate out the substrate degrees of freedom. For a frequency-independent tunneling amplitude 
(c.f. Eq.(\ref{hbsub})) the effects of the substrate can be parametrized by a constant level-broadening 
parameter $\G=\p\rho\zeta^2$, which measures the rate at which electrons in the bottom layer escape into the 
substrate. Then the retarded and Keldysh self-energies which renormalize the bottom Green's functions are given 
by
\begin{align}
\label{sigmar}
\S_k^R &=-i\G =(\S_k^A)^*, \\
\label{sigmak}
\S_k^K &=-2i\G[1-2g(\w)].
\end{align}
Here, $g(\w)$ is the Fermi-Dirac distribution of the substrate, where $\w$ is measured with respect to the
chemical potential $\mi_s$ of the substrate. We note here that because of our particular system geometry, 
the scattering into the bath dominates over electron-paramagnon scattering, which leads to considerable 
analytical simplifications. For instance, in Ref. \onlinecite{curr}, the geometry is that of a thin strip 
which is coupled to a bath only at its boundary. In that case, electron-paramagnon scattering is crucial and 
one must calculate the electron and paramagnon distributions self-consistently. We also note that coupling
the entire lower layer (and not only its boundaries) to an infinite bath allows one to assume that the Joule
heat generated by the electric field is efficiently carried away into the bath.

We now consider the effect of the nonequilibrium drive. At this point we specify the magnitude of the
electric field that will be considered in this work. We will be comparing the electric field $E$ to 
$\Gamma$, the escape rate of electrons from the bottom layer to the substrate, or rather to the corresponding 
time scale $\tau=1/2\Gamma$. We introduce a quantity which will be important in the remainder of this work: 
the effective temperature scale set by the nonequilibrium field, $T_{\rm eff}$. This quantity parametrizes the
consequences of the nonequilibrium-induced decoherence. In this system, where a uniform current flows in the 
inplane direction, $T_{\rm eff}$ is given by\cite{aditi2,curr}
\beq
\Te=eEv_F\tau.
\eeq 
We will limit ourselves to weak fields: $\Te/E_F\ll 1$ and $\tau\Te\ll 1$, where $E_F$ is the Fermi
energy of the bottom layer. Restricting ourselves to this regime and using the canonical
momentum, the retarded and advanced Green's functions take their equilibrium form\cite{aditi2}:
\beq
G^{R(A)}_{bk}=\frac{1}{\w-\xi^b_{\b k}\pm i\G}. 
\label{grab}
\eeq
Here, $\xi^b_{\b k}=\e_{\b k}-\mi_b$. The nonequilibrium Keldysh Green's function can be obtained 
using the linearized quantum Boltzmann equation\cite{mahan,rammer} (QBE) for the
lesser Green's function,
\beq
\label{qbe}
e{\b E}\cdot\round{\nabla_{\b k}+\b{v_k}\pd_\w}G^<_{bk}=\S^<_{bk}A_{bk}-2\G G^<_{bk}.
\eeq
With the usual parametrization, the lesser Green's function can be written as
\beq
G^<_{bk}=if_{bk}A_{bk},
\eeq
where $f_{bk}$ is the distribution function, and $A_{bk}$ the spectral function. With this parametrization
one obtains a QBE for the distribution function,
\beq
\label{QBEb}
e{\b E}\cdot\round{\nabla_{\b k}+\b{v_k}\pd_\w}f_{bk}=2\G[-f_{bk}+g(\w)].
\eeq
Following Ref. \onlinecite{aditi2}, we simplify Eq.(\ref{QBEb}) by dropping the gradient term in the left
hand side, which is justified in the weak-field limit. One then is required to solve the following QBE:
\beq
\label{qbe2}
e{\b E}\cdot\b{v_k}\pd_\w f_{bk}=2\G[-f_{bk}+g(\w)].
\eeq
The Keldysh Green's function is then obtained by solving Eq.(\ref{qbe2}) for the distribution function 
and substituting this into the standard formula,
\beq
G^K_{bk}=[1-2f_{bk}](G^R_{bk}-G^A_{bk}).
\label{gkbaditi}
\eeq
At zero temperature, the nonequilibrium electron distribution can be easily obtained and is given by
\beq
\label{aditif}
f_{bk}=\thi(-\w)+\frac{1}{2} [\sgn(\w)
+\sgn(\b E\cdot \b{v_k})]e^{-\frac{\abs{\w}}{\abs{e{\b E}\cdot\b{v_k}\tau}}},
\eeq
where we have used the fact that $g(\w)=\theta(-\w)$ at $T=0$ ($\theta$ being the Heaviside function). 
We see from Eqs.~(\ref{qbe2}) and (\ref{aditif}) that in equilibrium ($\b E = 0$), the bottom electron 
distribution reduces to the distribution of the substrate electrons signifying the state of 
thermal and chemical equilibrium between the two systems. 
We will hereafter assume for simplicity $\mi_t=\mi_b=\mi_s=\mi$. We therefore drop the 
superscript $t$ and $b$ from the top and bottom dispersions and assume that both 
dispersions and frequency $\w$ are measured with respect to the common chemical 
potential $\mi$.

We note here that the linear-response approach to obtaining the electron distribution function is 
justified here as long as the shift of the Fermi surface in momentum space ($eE\tau$) is much less than the 
Fermi wavevector. Indeed, we will find, as in similar previous works\cite{aditi1,aditi2}, 
that the nonequilibrium field is a 
relevant perturbation and grows under the renormalization group flow. In this work, we restrict our flow up 
to the scale where $eE\tau\sim k_F$ so that Eqs.(\ref{gkbaditi}) and (\ref{aditif}) remain valid.

\subsection{Interactions}
The interactions are given by
\beq
\label{interactingS}
\mathcal{S}^\ka_{\rm int}= \int d^3x [JS_b^{z,\ka}S_t^{z,\ka} + 
\frac{U}{2}(S_b^{z,\ka})^2 + \frac{U}{2}(S_t^{z,\ka})^2].
\eeq
The interactions are decoupled by two one-component bosonic fields, $m_b^\ka(x)$ and $m_t^\ka(x)$, 
which physically represent the magnetization in the bottom and top layers, 
respectively. The decoupling leads to
\begin{multline}
\mathcal{S}^\ka_{\rm int}=-U\int d^3x \left[\frac{1}{2}(m_b^\ka)^2+\frac{j}{2}m_b^\ka m_t^\ka\right. \\ -
\left.(m_b^\ka+jm_t^\ka)S_b^{z,\ka}\right] +[b\leftrightarrow t],
\label{SRT}
\end{multline}
where $j=J/U$ is the interlayer exchange coupling normalized by the intralayer exchange interaction.
At this point, it is useful to introduce a new basis for the bosonic fields:
\beq
\label{newbasis}
\cvec{\f_b^\ka}{\f_t^\ka}=\mat{1}{j}{j}{1}\cvec{m_b^\ka}{m_t^\ka}
\eeq
The transformation is well-defined given $j^2\neq 1$. The $\phi$ fields reduce to the physical
layer magnetizations when the interlayer coupling $J\rightarrow 0$. Both bases give an equivalent 
description of the problem but the analysis is simplified in the $\f$ basis because $\f_b$ ($\f_t$) 
couples directly only to $b$ ($t$) fermions, as can be seen from the third term in \req{SRT}. 
Integrating out the fermions will not generate additional couplings
between the bosons. In this new basis \req{SRT} takes the form
\begin{multline}
\mathcal{S}^\ka_{\rm int}=-U\int d^3x \frac{1}{1-j^2}\left[\frac{1}{2}(\f_b^\ka)^2+\frac{1}{2}(\f_t^\ka)^2
-j\f_b^\ka\f_t^\ka\right] \\ -\f_b^\ka S_b^{z,\ka}-\f_t^\ka S_t^{z,\ka}. 
\end{multline}
The final Keldysh action for the interactions is obtained after performing a change of basis in Keldysh space.
The corresponding transformation for the bosons reads
\beq
\f^{\rm cl}=\frac{\f^++\f^-}{\sqrt{2}},\quad \f^{\rm q}=\frac{\f^+-\f^-}{\sqrt{2}},
\eeq
with the analogous transformation for the conjugated fields. ``cl" and ``q" stand for classical and quantum
Keldysh components, respectively. The resulting action then reads
\begin{multline}
\label{SI}
\mathcal{S}^\ka_{\rm int} = 
\int_{k,q}\sum_{\s,\al}\frac{\s}{\sqrt{2}}\rvec{\bar{c}^1_{\al k \s}}{\bar{c}_{\al k\s}^2}
 \mat{ \f_\al^{{\rm cl}} }{ \f_\al^{\rm q} }{ \f_\al^{\rm q} }{ \f_\al^{\rm cl} }_{k-q}
\cvec{c^1_{\al q\s}}{c^2_{\al q\s}} \\
-\frac{1}{2\t U}\int_q\sum_{\al,\be} \left[ \f_{\al,-q}^{\rm cl} \round{\de_{\al\be}
-j (\tau_x)_{\al\be}} \f_{\be q}^{\rm q} + {\rm c.c.} \right],
\end{multline}
where $\t U=U(1-j^2)$. $\al$ and $\be$ here are summed over $b$ and $t$, and
$q=(\W,{\b q})$ labels bosonic frequency and momentum.

\subsection{Integrating out fermions}
We now integrate out all remaining fermionic modes to obtain an effective bosonic theory in the $\f$ basis.
Here, we follow the standard procedure and expand the resulting action up to quartic order 
in the bosonic fields. This then yields
\beq
\mathcal{S}_{\rm eff}=\mathcal{S}^{(2)}_{\rm eff}+\mathcal{S}^{(4)}_{\rm eff},
\eeq
where the gaussian part reads 
\begin{multline}
\label{sgausseff}
\mathcal{S}_{\rm eff}^{(2)} = 
-\int_{q}\left\{ \Phi_{b q}^\dagger \left[(\t U\nu)^{-1} \hat\tau_x +\hat\Pi_{bq}\right]
\Phi_{b q}\right. \\ \left.- \Phi_{b q}^\dagger (\e\hat\tau_x) \Phi_{t q} + (b\leftrightarrow t)\right\},
\end{multline}
where $\e=j(\t U\nu)^{-1}$, $\nu=m/\p$ is the two-dimensional density of states at the Fermi
energy and $\hat\tau_x$ is a Pauli matrix in Keldysh space. 
The two-component vector $\Phi_{\al}^T=\rvec{\f^{\rm cl}_\al}{\f^{\rm q}_\al}$ denotes the structure 
of the fields in Keldysh space. The hat above the polarization function, $\hat\Pi_{\al q}$, denotes its matrix nature 
with the usual Keldysh causality structure\cite{kamenev,rammer},
\beq
\label{picausality}
\hat\Pi_{\al q}=\mat{0}{\Pi^A_{\al q}}{\Pi^R_{\al q}}{\Pi^K_{\al q}}.
\eeq
Here, the retarded and Keldysh polarization functions are defined by
\Beq
\label{pirf}
\Pi^R_{\al q} &=& \frac{-i}{\nu}\int_k[G^R_{\al, k+q}G^K_{\al k}+G^K_{\al, k+q}G^A_{\al k}]; \\
\label{pikf}
\Pi^K_{\al q} &=& \frac{-i}{\nu}\int_k [G^K_{\al, k+q} G^K_{\al k}
+G^R_{\al, k+q}G^A_{\al k}+G^A_{\al, k+q} G^R_{\al k}],
\Eeq
and $\Pi^A_{\al q}=(\Pi^R_{\al q})^*$.
The quartic terms read
\beq
\label{s4eff}
\mathcal{S}_{\rm eff}^{(4)} = \int d^3x \sum_{\al,n=1}^{n=4}u_n^\al [\f^{\rm cl}_{\al}(x)]^{4-n}
[\f^{\rm q}_{\al}(x)]^n.
\eeq
Note that as advertised above, the quartic part of the action does not
couple the bosonic fields. As a matter of fact, in the $\phi$-basis, the \emph{only}
mixing appears in the quadratic part of the action.
We now turn to the evaluation of the polarization functions, which will reveal the dynamics
of the bosonic fields as well as the effects of the non``equilibrium drive on the latter.

Integrating out soft fermionic modes is now known to yield non-analytic corrections to
the static spin susceptibility and invalidate the standard Landau-Ginzburg-Wilson description for itinerant
ferromagnets in relevant dimensions\cite{BKV}. However, these non-analyticities are also known to cancel
for the case of ordering field with Ising symmetry\cite{chubukov}. 

\subsection{Polarization functions}
We begin with the evaluation of $\hat\Pi_{tq}$, which is computed using equilibrium
electron propagators of the top layer. The retarded component can then be obtained by analytic
continuation of the standard Matsubara polarization function used for a clean 
itinerant ferromagnet\cite{hertz,millis,moriya} yielding
\beq
\label{pirt}
\Pi^R_{tq}=-1+|{\b q}|^2-i\frac{\W}{v_F|{\b q}|}.
\eeq
The Keldysh polarization function for the top layer can be straightforwardly obtained
using the fluctuation-dissipation theorem. At $T=0$, it reads
\beq
\label{pikt}
\Pi^K_{tq}=-2i\frac{|\W|}{v_F|{\b q}|}.
\eeq

The polarization functions, $\hat\Pi_{bq}$, have been obtained elsewhere\cite{aditi2} and the 
detailed calculations here will be relegated to Appendix \ref{bottomPF}. Here we state the results. 
The retarded polarization function for $|{\b q}|<(v_F\tau)^{-1}$ and $T_{\rm eff}\tau\ll 1$ is given by
\beq
\label{pirbexpp}
\Pi^R_{bq}=-1+|{\b q}|^2-i\tau\round{\W-\b v_d\cdot{\b q}}.
\eeq
We note here that the condition $|{\b q}|<(v_F\tau)^{-1}$ was crucial in obtaining the $z=2$ dynamical term
above with a constant Landau damping coefficient. In the limit where $|{\b q}|>(v_F\tau)^{-1}$,
an expansion with respect to $(|{\b q}|v_F\tau)^{-1}$ becomes possible and the dynamical term for
the bottom layer gets modified to $i\W/v_F|{\b q}|$. We shall focus on excitation energies less than
$\G$ and momenta less than $\G/v_F$ where nonconservation due to escape into the leads is dominant.

The parity-breaking drift term contains the drift velocity which reads
\beq
\b v_d=\frac{e{\b E}\tau}{m}.
\eeq
The Keldysh polarization function contains information about decoherence arising from the noise 
in the current, and is given by
\beq
\Pi^K_{bq}=-2i\abs{\W}\tau\square{1+I\round{\frac{T_{\rm eff}}{\abs{\W}}}},
\label{pikbexp}
\eeq
where
\beq
I(x):=x\int\frac{d\thi}{2\p}\abs{\cos\thi}e^{-\frac{1}{x\abs{\cos\thi}}}.
\eeq
Absence of drift in the Keldysh polarization indicates that the noise does not drift in the presence of 
the current in the bottom layer. 

Once these polarization functions are inserted into Eq.(33) one obtains a theory of two coupled 
bosonic modes, one of which ($\Phi_b$) is characterized by a dynamical critical exponent $z=2$
and the other ($\Phi_t$) by $z=3$. In the decoupled limit ($\epsilon\rightarrow 0$), the former 
fluctuations are non-conserved in nature because the bottom layer is open to the external bath. On the other 
hand, the top layer does not couple to any bath and is thus characterized by conserved fluctuations.

\section{Preliminary considerations}
\label{precons}
\subsection{Mean-field analysis}
\label{MF}
We begin with a mean-field analysis of the effective bosonic action (Eqs.(\ref{sgausseff}),(\ref{s4eff})) 
in order to understand how
a finite interlayer interaction and nonequilibrium drive affect the underlying Stoner condition. 
Assuming a mean-field ferromagnetic state, whose solution is both static and uniform (i.e.
$\W\rightarrow 0$ and ${\b q}\rightarrow {\b 0}$), we
may write down the following Landau-Ginzburg theory for the bilayer system,
\beq
\label{LGTheory}
\mathcal{S}_{\rm LG}=\de(\f_b^2+\f_t^2)-2\e\f_b\f_t +u_b\f_b^4+u_t\f_t^4.
\eeq
Here, $\de=1/(\t U\nu) - 1$; we have set the area to unity. It is clear from Eq.(\ref{LGTheory}) that the 
interlayer coupling $j$ gives rise to a renormalization of the Stoner criterion. However, the 
(parity-breaking) electric field does not give an additional renormalization as the relevant corrections 
vanish for $\W\rightarrow 0$ and ${\b q}\rightarrow {\b 0}$.\cite{stoner} 

\begin{figure}
\begin{center}
\includegraphics[scale=0.5]{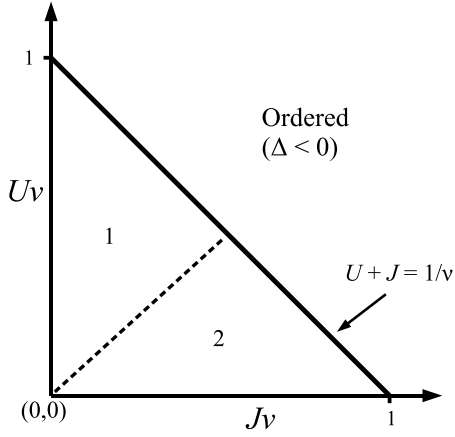}
\caption{\label{fig:pd} Mean field phase diagram. The solid line separates the ordered ($\Delta<0$) and 
disordered ($\Delta>0$) phases. The dashed line corresponds to $U=J$; in region 1 (2), $\epsilon>0$ ($<0$).  In this work, we concentrate 
on region 1 of the disordered phase, near the transition line. }
\end{center}
\end{figure}

When $\e=0$, i.e. $j=0$, the two fields decouple and the usual Stoner condition for the instability (i.e. $1/U_c=\nu$) 
is regained for each of the two layers.  However, for $\e>0$, the Stoner condition for a critical $U$ is shifted.
To see this, we begin with the saddle-point conditions for the theory,
\Beq
&&\de\f_b-\e\f_t+2u_b\f_b^3=0,\\
&&\de\f_t-\e\f_b+2u_t\f_t^3=0.
\Eeq
In general, $u_b\ne u_t$, because these constants microscopically arise from particle-hole fluctuations, 
and the fermions in the bottom and top layers possess different dynamics. 
We see that in the mean-field equation for a given layer, the complimentary layer enters like a magnetic 
field that tends to align both order parameters. Decoupling the equations leads to $0=(\de^2/\e-\e)\f_\al+\frac{\de}{\e}(u_\al+\frac{\de^2}{\e^2}
u_{\bar\al})\f_\al^3+\frac{3\de^2u_\al u_{\bar\al}}{\e^3}\f_\al^5+\frac{3\de u_\al^2u_{\bar\al}}{\e^3}
\f_\al^7+\frac{u_\al^3u_{\bar\al}}
{\e^3}\f_\al^9$, where $\bar{\al}$ is the label complimentary to $\al$. We see that for $\de>\e$ the system is 
disordered and that the transition occurs at $\de=\e$ as one increases $U$. The
modified Stoner condition for the instability is then given by
\beq
\label{newstoner}
U_c=\frac{1}{\nu}-J.
\eeq
We see that the critical $U$ is reduced from the one corresponding to the single layer Stoner
criterion because the interlayer interaction promotes the ferromagnetic state. The considerations above 
lead us to introduce
the tuning parameter for the transition $\Delta=\delta-\epsilon$. The mean field phase diagram is shown in Fig. \ref{fig:pd}.  
Throughout this work, we will be considering properties of the system in the vicinity of the transition and
in the disordered region of the phase diagram where $\Delta>0$, i.e. $\de>\e$. In addition, we will be assuming
 $\de\gtrsim\e>0$, and $\Delta \ll \de,\,\e$, which corresponds to a region close to the transition
line in region 1 of Fig.~\ref{fig:pd}.

\subsection{Fixed point}
\label{fp}

The effective bosonic action (Eqs.(\ref{sgausseff}),(\ref{s4eff})) consists of two 
coupled modes, $\f_b$ and $\f_t$, that are characterized by different dynamical critical exponents. In the 
decoupled limit ($\e\rightarrow 0$), the $\f_b$ spin fluctuations are non-conserved in nature because the 
bottom layer is open to the substrate, and this gives rise to its $z=2$ dynamics. $\f_t$ fluctuations, on the 
other hand, are conserved with $z=3$ dynamics because the top layer is not subject to particle exchange with 
an external bath. For finite $\e$, these modes are coupled at the gaussian level. As such, the model
exhibits multi-scale quantum criticality. The phenomenon has been addressed in the context of the Pomeranchuk 
(or nematic) instability in two spatial dimensions\cite{oganesyan,koln}, and in the study of quantum phase
transitions that spontaneously develop ferromagnetic order in a helical Fermi liquid\cite{cenke}, the latter
describing the low-energy physics of surface states of a three dimensional topological insulator.
In both situations, the low-energy bosonic theory consists of two modes, one longitudinal and the other 
transverse, with different dynamical critical exponents. The lowest order coupling appears at the quadratic level,
i.e. $\phi_\perp^2\phi_\parallel^2$, in contrast with the linear coupling present in our situation. In both 
the Pomeranchuk and helical liquid scenarios, the longitudinal mode is Landau-damped with $z=3$ and the transverse 
one is ballistic and undamped with $z=2$. In Ref. \onlinecite{koln}, 
it was argued that the low-energy behaviour of the system at $T=0$ is governed by the undamped $z=2$ 
mode since it possesses the smaller effective dimension. This claim is supported by the fact that 
renormalizations due to fluctuations to the mass and the vertex are dominated in the infrared by the 
$z=2$ mode. 

\begin{figure}[t]
\begin{center}
\includegraphics[scale=0.8]{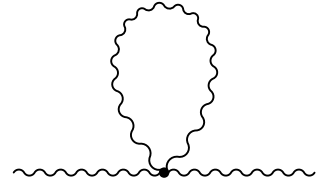}
\caption{\label{fig:tadpole} Tadpole diagram for the mass renormalization. The curly lines correspond 
to magnetization fluctuations $\phi_\al$ belonging to the same sector, $b$ or $t$. As noted before, 
there's no mixing at the quartic level in the $\phi$ basis.}
\end{center}
\end{figure}

To carefully determine the nature of the equilibrium fixed point at $T=0$, let us consider the 
renormalization to the mass in perturbation theory. 
We begin with our gaussian action in the Matsubara form
\beq
\label{eqaction}
\mathcal{S}^{(2)}_{\rm eq}=\int_q \rvec{\f_{bq}^*}{\f_{tq}^*}
\mat{\cL^{\rm eq}_{bq}}{-\e}{-\e}{\cL^{\rm eq}_{tq}}\cvec{\f_{bq}}{\f_{tq}},
\eeq
where $\cL^{\rm eq}_{bq}=\de+|{\b q}|^2+|\W|\tau$ and $\cL^{\rm eq}_{tq}=\de+|{\b q}|^2+|\W|/v_F|{\b q}|$
are the equilibrium inverse susceptibilities.
The propagators for the modes, obtained by inverting the $2\times 2$ inverse susceptibility matrix, are
$\langle\f_b(q)\f_b^*(q)\rangle=\cL^{\rm eq}_{tq}/D^{\rm eq}_q$, 
$\langle\f_t(q)\f_t^*(q)\rangle=\cL^{\rm eq}_{bq}/D^{\rm eq}_q$,
and $\langle\f_b(q)\f_t^*(q)\rangle=\langle\f_t(q)\f_b^*(q)\rangle=\e/D^{\rm eq}_q$, where 
$D^{\rm eq}_q=\cL^{\rm eq}_{bq}\cL^{\rm eq}_{tq}-\e^2$ is the determinant of the matrix. While the off-diagonal 
mass, $\e$, does not gain corrections from quartic fluctuations, the diagonal mass, $\de$, does gain 
renormalizations corresponding to the usual tadpole diagram of Fig.\ref{fig:tadpole}.
For the $\f_b$ sector, this is given by 
\beq
\label{mr1}
\De\de_b\propto\int\frac{d^2q}{(2\p)^2}\frac{d\W}{2\p}\frac{\de +|{\b q}|^2+\frac{|\W|}{v_F|{\b q}|}}
{D^{\rm eq}_q}.
\eeq
The contribution of the $z=3$ fluctuations near the critical point can be estimated by setting
$\W\sim |{\b q}|^3$. Then Eq.(\ref{mr1}) essentially reduces to
\beq
\left.\De\de_b\right|_{z=3}\sim\int \frac{d^5q}{2\De+ |{\b q}|^2}\sim \De^{3/2},
\eeq
where $\De=\de-\e$ is the distance to the critical point, and we have assumed $\de\approx\e$. 
For $z=2$ fluctuations, $\W\sim |{\b q}|^2$, and the contribution can be estimated as
\beq
\left.\De\de_b\right|_{z=2}\sim\int \frac{d^4q}{2 \De+ |{\b q}|}\sim \De^{3}.
\eeq
In the vicinity of the critical point (as $\De\rightarrow 0$), the $z=3$ fluctuations are expected
to dominate. A similar calculation can be carried out for the $\f_t$ sector leading to the same conclusion.

The key difference here from the analysis in Ref. \onlinecite{koln} is that the modes are coupled 
at the gaussian level, and that the two dynamical terms enter together in all of the propagators. It is then
clear that close enough to the critical point (where the characteristic length scale of the bosonic 
fluctuations $\xi>v_F\tau$), the conserved dynamical term ($\W/v_F|{\b q}|$) will dominate over the
non-conserved counterpart ($\W\tau$), and the low-energy long-wavelength theory will be governed
by $z=3$ dynamics. We now study the renormalization group flow in the vicinity of this
$z=3$ fixed point.

\section{Renormalization Group Analysis}
\label{RGA}
To aid with the analysis we begin with the effective gaussian action (c.f. Eq.(\ref{sgausseff})),
\begin{multline}
\mathcal{S}^{(2)}_{\rm eff}=-\int_q^\La \rvec{\Phi_{bq}^\dag}{\Phi_{tq}^\dag} 
\mat{\hat\cL_b}{-\e\hat\tau_x}{-\e\hat\tau_x}{\hat\cL_{t}} \cvec{\Phi_{bq}}{\Phi_{tq}},
\label{sgausseffrg}
\end{multline}
where $\hat\cL_\al=(\t U\nu)^{-1}\hat\tau_x+\hat\Pi_{\al q}$. The explicit expressions for the matrix
elements read
\Beq
\cL_b^R&=&\de+|{\b q}|^2+i\b v_d\cdot{\b q}\tau-i\W\tau,\\
\cL_b^K&=&-2i|\W|\tau\square{1+I\round{\frac{T_{\rm eff}}{|\W|}}},
\Eeq
and
\Beq
\cL_t^R&=&\de+|{\b q}|^2-i\frac{\W}{v_F|{\b q}|},\\
\cL_t^K&=&-2i\frac{|\W|}{v_F|{\b q}|}.
\Eeq
The quartic action was given in Eq.(\ref{s4eff}). 

\subsection{Flow equations out of equilibrium}

We study the flow in the vicinity of the $z=3$ gaussian fixed point with $T_{\rm eff}=\De=u^\al_n=0$. 
Here, $T=0$ throughout. At this fixed point the most relevant dynamical term 
(i.e. $\W/v_F|{\b q}|$) remains marginal. In Eq.(\ref{sgausseffrg}), we have defined our momentum cutoff $\La$,
which in turn defines an associated cutoff energy scale $v_F\La$. All energy quantities will be considered 
to be in units of this scale. The bare parameters are taken to be 
$\De,\, T_{\rm eff},\, u_n^\al,\,\de,\,\e\ll 1$ with $\de\sim\e\gg\De$. 
We begin by integrating out fluctuations whose modes reside 
in the shell $\La/b\le q \le \La$. Here, $b>1$ is the scaling variable. In order to preserve Keldysh causality, we 
integrate over all frequencies (i.e. $-\infty<\W<\infty$) at every mode elimination step. We next 
rescale momentum, $q\rightarrow q'/b$, to restore the cutoff back to $\La$. At the
$z=3$ fixed point, the frequency scales as $\W\rightarrow\W'/b^3$ and this keeps the 
coefficients of $|{\b q}|^2$ and $\W/v_F|{\b q}|$ terms invariant under the flow provided that the fields 
scale as $\f_{\al q}^{\rm cl,q}\rightarrow b^{7/2}[\f_{\al q}^{\rm cl,q}]'$. The $(T_{\rm eff}/\W)$-scaling 
present in the Keldysh components suggests that 
$T_{\rm eff}$ is relevant at the fixed point and scales with dimension of 3, 
i.e. $T_{\rm eff}\rightarrow T'_{\rm eff}/b^3$. The non-conserved dynamical term is irrelevant
at the fixed point, and this is reflected in the scaling $\tau\rightarrow\tau' b$. Then, the requirement to 
maintain the drift term invariant under the flow dictates the scaling of the drift velocity as 
$v_d\rightarrow v_d'/b^2$. Note that the drift term contains the combination $\tau v_d$, which has scaling
dimension 1, just as in Ref. \onlinecite{aditi2}.
Both the diagonal ($\de$) and the off-diagonal ($\e$) mass terms scale with 
the usual dimension of 2. Finally, at the $z=3$ fixed point, all quartic coupling constants are irrelevant with 
$[u^\al_n]=-1$. Therefore, in the standard lore\cite{hertz,millis}, the order parameter self-interactions are not 
expected to play a significant role in the effective low-energy theory. 

To summarize our scaling analysis, we write down the set of RG equations to lowest (linear) order in the 
quartic coupling,
\Beq
\frac{dT_{\rm eff}(b)}{d\ln b}&=&3T_{\rm eff}(b),\\
\frac{d\e(b)}{d\ln b}&=&2\e(b),\\
\frac{dv_d(b)}{d\ln b}&=&2v_d(b),\\
\frac{d\tau(b)}{d\ln b}&=&-\tau(b),\\
\frac{du_n^\al(b)}{d\ln b}&=&-u_n^\al(b)+\mathcal{O}([u_n^\al]^2),\\
\label{deltaflow}
\frac{d\de_\al(b)}{d\ln b}&=&2\de_\al(b)+3u_1^\al(b) f_\al(T_{\rm eff}(b),\tau(b)).
\Eeq
We will distinguish renormalized parameters from the bare ones by explicitly writing the $b$-dependence 
for the former. In Eq.(\ref{deltaflow}), $\al$ labels $b$ and $t$. Indeed, the bare values satisfy $\de_b=\de_t
=\de$; the label denotes the fact that these mass terms are subject to different 
renormalizations from fluctuations. The function, $f_\al(T_{\rm eff}(b),\tau(b))$ is given by
\beq
\label{ffunc}
f_\al(T_{\rm eff}(b),\tau(b))=iK_2\int\frac{d\W}{2\p}D^K_\al(1,\W),
\eeq
where $K_2=\int\frac{d^2\b q}{(2\p)^2}\de(q-1)=1/2\p$, we have set the momentum cutoff $\La=1$, and
\beq
\label{dk}
D^K_{\al p}=\frac{\cL^K_{\al p}|\cL^R_{\bar\al p}|^2+\e^2\cL^K_{\bar\al p}}{|\cL^R_{b p}
\cL^R_{t p}-\e^2|^2}
\eeq
is the Keldysh Green's function for the mode $\phi_\al$, it was obtained by inverting the inverse susceptibility
matrix of Eq.~(\ref{sgausseffrg}). Solving these equations, we arrive at
\Beq
T_{\rm eff}(b)&=&T_{\rm eff}b^3\\
\e(b)&=&\e b^2,\\
v_d(b)&=&v_d b^2,\\
\tau(b)&=&\tau b^{-1},\\
u_n^\al(b)&=&u_n^\al b^{-1},\\
\label{desolneq}
\de_\al(b)&=&b^2[\de+3u_1^\al\int_0^{\ln b}dxe^{-3x}f_\al(T_{\rm eff}e^{3x},\tau e^{-x})].
\Eeq

\subsection{Flow equations in equilibrium}
To elucidate the nonequilibrium problem at later stages, we also consider flow equations for an 
equilibrium situation in which the entire system is at temperature $T$. Here, the electric field is set
to zero, thus $T_{\rm eff}=0$. 
In this case, the gaussian action, Eq.(\ref{sgausseffrg}), becomes
\Beq
\label{polfunceq1}
\cL_b^R&=&\de+|{\b q}|^2-i\W\tau,\\
\cL_b^K&=&-2i\W\tau\coth\round{\frac{\W}{2T}},
\Eeq
and
\Beq
\cL_t^R&=&\de+|{\b q}|^2-i\frac{\W}{v_F|{\b q}|},\\
\label{polfunceq2}
\cL_t^K&=&-2i\frac{\W}{v_F|{\b q}|}\coth\round{\frac{\W}{2T}}.
\Eeq
An analogous renormalization group analysis as in the nonequilibrium case can be applied here, and the 
solutions to the corresponding flow equations become
\Beq
T(b)&=&T b^3\\
\e(b)&=&\e  b^2,\\
\tau(b)&=&\tau b^{-1},\\
u_n^\al(b)&=&u_n^\al b^{-1},\\
\label{desoleq}
\de_\al(b)&=&b^2[\de +3u_1^\al\int_0^{\ln b}dxe^{-3x}f_\al(T e^{3x},\tau e^{-x})].
\Eeq
The $f_\al$-functions are still given formally by Eqs.(\ref{ffunc}),(\ref{dk}) with the replacement 
$T_{\rm eff}\rightarrow T$, and the equilibrium inverse susceptibilities, Eqs.(\ref{polfunceq1})-(\ref{polfunceq2}), 
must be used.

\section{Correlation length}
\label{CLs}
In the current bosonic basis, there are three two point correlation functions one can study, each with its own 
associated mass scale, which we define as 
\beq
\Delta_{\al\be}^{-1}=\langle \phi_{\al q}\phi_{\be q}^*\rangle |_{q=0},
\eeq
where $q$ is a frequency-momentum 3-vector. In the above equation we have omitted the Keldysh indices
because one can use an equilibrium formalism to define the mass scales as they
don't depend on the nonequilibrium drive. Note also that $\Delta_{bt}=\Delta_{tb}$. Within the gaussian theory, the bare mass scales read
\Beq
\label{xi11}
\De_{bb}&=&\De_{tt}=\frac{\de^2-\e^2}{\de},\\
\label{xi12}
\De_{bt}&=&\frac{\de^2-\e^2}{\e}.
\Eeq
In the relevant limit, $\de>0$, $\e>0$, and $\de\gtrsim\e\gg \Delta$, Eqs.(\ref{xi11}),(\ref{xi12}) become
\beq
\De_{bb}=\De_{tt}\approx\De_{bt}\approx 2\De.
\eeq
Here, the mass scales were computed to lowest order in $\De/\de\ll1$ and $\De/\e\ll1$. One sees
that they all vanish as the transition is approached ($\De\rightarrow0$). This allows one to then
define the physical correlation length of the system, $\xi$, via $\xi^{-2}:=\De$.
We now compute $\xi$ both in and out of equilibrium focusing on the corrections arising from either
$T$ or $T_{\rm eff}$.

\subsection{In equilibrium}
We begin by considering the correlation length in equilibrium. We define the quantum-to-classical 
crossover line for the system, and compute the leading temperature correction to the correlation length 
in the quantum critical regime.

\subsubsection{Quantum disordered regime}
In this regime, the $f_\al$-functions are computed at $T=0$. This gives a temperature independent shift 
proportional to $u_1^\al$ to the two diagonal masses. We then have
\begin{align}
\de_\al(b)&=b^2[\de +3u_1^\al\int_0^\infty dxe^{-3x}f_\al(0,\tau e^{-x})],\notag \\
&:= b^2(\de +\De\de_\al^0).
\end{align}
The correlation length in this regime is then given by
\beq	
\label{CLqeq}
\xi^{-2}\approx \De+\De\de_b^0+\De\de_t^0=:r,
\eeq
where we have computed to lowest order in $\De/\de \ll 1$ and $\De\de_\al^0/\de \ll 1$. To find the condition 
on $T $ for the occurrence of the regime, we impose that $\Delta(b')=1$ while $T(b')< 1$. Recall that
$\Delta(b)\approx\Delta_{bb}(b)\approx\Delta_{tt}(b)\approx\Delta_{bt}(b)$. This then translates
to a condition on the bare quantities,
\beq
\label{qtoceq}
T < r^{3/2}.
\eeq

\subsubsection{Quantum critical regime}
\label{CLeq}
When the inequality in Eq.(\ref{qtoceq}) is reversed, the characteristic energy scale of the fluctuations,
$1/\xi^z$,  become smaller than temperature and the fluctuations become classical.
Here, we compute the leading temperature correction to the correlation length at 
scale $b^*$, where $T(b^*)=1$. Since $\de \sim\e \gg r \approx T^{2/3}$,
both $\de$ and $\e$ become much larger than 1 at some stage of the flow. We define the scale
$b_1$ as $\de(b_1)\sim\e(b_1)=1$. Then the integral in Eq.(\ref{desoleq}) must be split into two regions: 
$1<b<b_1$, and the other $b_1<b<b^*$. In the former region, the integral
can be computed assuming $\de\sim\e\ll 1$ inside Eq.(\ref{dk}). In the latter region, the integral is computed
assuming $\de\sim\e\gg 1$. The detailed derivation will be relegated to Appendix \ref{CLapp}; 
here we state the results. We define
\begin{multline}
\De\de_\al:=3u_1^\al\int_0^{\ln b^*}dx\,e^{-3x}
[f_\al(T e^{3x},\tau e^{-x})-f_\al(0,\tau e^{-x})].
\end{multline}
Here, one integrates up to $b^*$ by splitting the integral as advertised. One then
arrives at the following temperature correction to leading order,
\beq
\De\de_\al=\frac{u_1^\al}{4\p^2}T \round{1+\frac{3\tau }{2}T^{1/3}}.
\eeq
The correlations length in the quantum critical regime is then given by
\beq
\label{CLstrongeq}
\xi^{-2}_{\rm eq}\approx r+\frac{u_1^b+u_1^t}{4\p^2}T\round{1 +\frac{3\tau }{2}T ^{1/3}}.
\eeq
The linear-temperature correction in Eq.(\ref{CLstrongeq}) is consistent with the corresponding correction
obtained in Ref. \onlinecite{millis} for the case $d=2$ and $z=3$. The $T^{4/3}$-correction, which is
proportional to $\tau $, is an additional correction that arises because of the $z=2$ dynamics present
at the $z=3$ fixed point. Both the leading $z=3$ correction and a sub-leading $z=2$ correction 
enter into the correction because both modes are subject to a common temperature $T$.

\subsection{Out of equilibrium}
\label{CLneq}
We now compute the leading nonequilibrium correction to the correlation length at scale $b^*$, 
where $T_{\rm eff}(b^*)=1$. We now assume $T =0$. Here, we only concentrate on the quantum critical regime 
because the correlation length in the quantum disordered regime is still given by Eq.(\ref{CLqeq}). Of course, the 
condition for the occurrence of this regime now reads $T_{\rm eff} < r^{3/2}$.

The leading correction due to the electric field can be computed by splitting the scaling regimes 
into $1<b<b_1$ and $b_1<b<b^*$ as in the equilibrium case. We then obtain,
\beq
\De\de_\al\approx\frac{3u_1^\al}{4\p^3}\tau T _{\rm eff}^{4/3}.
\eeq
The correlations length in the quantum critical regime is then given by
\beq
\label{CLstrongneq}
\xi^{-2}_{\rm neq}\approx r+3\frac{u_1^b+u_1^t}{4\p^3}\tau T _{\rm eff}^{4/3}.
\eeq
Once again, the detailed calculations are presented in Appendix \ref{CLapp}. 
In contrast to the equilibrium result, Eq.(\ref{CLstrongeq}), the correlation length gains a correction of
order $T_{\rm eff}^{4/3}$ only. The difference arises because in the decoupled limit the two layers 
are at different effective temperatures. This can be readily checked by comparing the two Keldysh polarization 
functions, Eqs.(\ref{pikt}) and (\ref{pikbexp}). One sees that the top layer is at zero effective temperature while 
the bottom possesses a finite $T_{\rm eff}$ because the field directly couples to the latter. Once
the interlayer coupling is restored both $\phi_b$ and $\phi_t$ sectors feel the effect of the field as it is evident 
from Eq.(\ref{CLstrongneq}). However, due to the absence of an effective temperature in the top Keldysh 
polarization function, a linear-$T_{\rm eff}$ correction does not arise in Eq.(\ref{CLstrongneq}).

Results in this section suggest certain similarities between the role of  
$T_{\rm eff}$ in the nonequilibrium case and temperature $T$ in the equilibrium case. Both $T_{\rm eff}$
and $T$ induce a quantum-to-classical crossover, across which the behaviour of critical fluctuations evolves 
from quantum to effectively classical. The crossover energy scale is defined in Eq.(\ref{qtoceq}) where $T$ can
also be replaced by $T_{\rm eff}$. Furthermore, Eq.(\ref{CLstrongneq}) implies that $T_{\rm eff}$ can cutoff the
divergence in the correlation length just as temperature $T$ can as shown in Eq.(\ref{CLstrongeq}). 
However, the lack of a linear-$T_{\rm eff}$ correction for the correlation length indicates that 
$T_{\rm eff}$ cannot be identified \textit{strictly} as an effective temperature of the system.

We note that in equilibrium two corrections that reflect the presence of $z=2$ and $z=3$ dynamics enter 
into the correlation length. The dominant contribution (i.e. linear-$T$ contribution) 
arises essentially from the $z=3$ dynamical terms in the regime of interest. In systems with multi-scale quantum criticality,
modes with different dynamics often nontrivially contribute to various thermodynamic quantities in different 
bare parameter regimes. In thermal equilibrium, interesting crossover behaviour in thermodynamic quantities 
in a system with multi-scale quantum criticality has been addressed in the context of Pomeranchuk 
instability\cite{koln} and the Kondo-Heisenberg model\cite{pepin}.

\subsection{Eigenmode analysis}
We have found that one must use caution in determining the mass scale which 
relates to the inverse square of the physical correlation length, for neither $\de$ nor $\e$ alone defines 
this scale. A sensible procedure would be to determine the eigenmodes of the system that diagonalize the 
gaussian action Eq.(\ref{sgausseff}), and directly read off the corresponding masses by observing the 
eigenvalues of the action. This analysis was carried out, and is presented in Appendix \ref{eigenbasisanal}.
We find that while one of the eigenmodes becomes critical at $\de=\e$, the other remains massive.
We then integrate out the massive mode and construct an effective critical theory in terms of a single
critical mode, $\la_q$. In the appendix, we carry out a renormalization group procedure on
this single-component critical theory and the corresponding $T$ and $T_{\rm eff}$ corrections to the correlation 
length of this field is computed. We find that the correlation length obtained in the eigenbasis gives identical 
results to those presented in Eqs.(\ref{CLstrongeq}),(\ref{CLstrongneq}).

\section{Order parameter dynamics}\label{langeq}
We now consider the dynamics obeyed by the critical fluctuations in the quantum critical regime 
where the theory is effectively classical and the quantum fluctuations can be integrated out from the
effective low-energy theory. The standard procedure to obtain the dynamics is outlined in 
Ref. \onlinecite{kamenev}, and has been applied in Refs.\onlinecite{aditi1,aditi2} in establishing 
nonequilibrium dynamical universality classes. In a completely different setting but in a related work, 
order parameter dynamics near phase transitions in a driven interacting bilayer lattice gas has also 
been addressed in Refs.\onlinecite{schmittmann1,schmittmann2}.

We begin by decoupling the terms quadratic in 
the quantum Keldysh fields by making use of a second Hubbard-Stratonovich transformation.
The decoupling field plays the role of a noise source for the classical fluctuations. Finally, 
we integrate out the quantum fluctuations and obtain a stochastic equation, or 
Langevin equation, for the classical fluctuations. We hereafter drop the ``cl" subscript
from the fields. All couplings will also be assumed to take their renormalized values.

In the $\f_{\al}$ basis, the system of coupled, linear Langevin equations is 
\beq
\mat{\cL_{bq}^R}{-\e}{-\e}{\cL_{tq}^R}\cvec{\f_{bq}}{\f_{tq}}=\cvec{\eta_{bq}}{\eta_{tq}},
\eeq
where $\eta_{\al q}$ are the Fourier components of the noise sources; their correlators are given by the Keldysh inverse susceptibilities
\beq
\langle\eta_{\al q}\eta_{\be q'} \rangle=\de_{\al\be}\de(q+q') i\Pi_{\al q}^K.
\eeq
One can obtain the noise-averaged correlators for the fluctuations by inverting the coefficient matrix:
\begin{align}
\langle \f_{bq}\f_{bq}^* \rangle &= \frac{|\cL_{tq}^R |^2 }{|D_q|^2}i\Pi_{bq}^K+ \frac{\e^2}{|D_q|^2}i\Pi_{tq}^K, \\
\langle \f_{tq}\f_{tq}^* \rangle &= \frac{|\cL_{bq}^R |^2 }{|D_q|^2}i\Pi_{tq}^K+ \frac{\e^2}{|D_q|^2}i\Pi_{bq}^K, \\
\langle\f_{tq}\f_{bq}^* \rangle &= \frac{\e}{|D_q|^2} \left[ \cL_{tq}^Ri\Pi_{bq}^K+ \cL_{bq}^A i\Pi_{tq}^K \right],
\end{align}
where $D_q=\cL_{bq}^R\cL_{tq}^R-\e^2$ is the determinant of the coefficient matrix. We mention here that for finite
interlayer coupling all bosonic correlators feel the presence of the parity-breaking drift term because they all contain $\cL_{bq}^R$.
One can repeat the same procedure for the critical eigenmode (see Appendix \ref{eigenbasisanal}), 
whose Langevin equation reads
\beq
\label{formalle}
\chi_R^{-1}\la_q=\eta_q,
\eeq
with the noise correlator $\ang{\eta_{q} \eta_{q'}} = i\chi_K^{-1}\delta(q+q')$.
From Appendix \ref{eigenbasisanal}, we have that
\begin{align}
\chi_R^{-1} &=\frac{1}{2}(\cL_{bq}^R+\cL_{tq}^R)-\e, \\
\chi_K^{-1} &=\frac{1}{2}(\Pi_{bq}^K+\Pi_{tq}^K).
\end{align}
The noise-averaged correlator for the eigenfield then reads
\beq
\langle\la_q\la_q^*\rangle= \frac{\frac{i}{2}
(\Pi_{bq}^K+\Pi_{tq}^K)}{|\frac{1}{2}(\cL_{bq}^R+\cL_{tq}^R)-\e |^2}.
\eeq
We now argue that the three correlators in the $\f$ basis and the
one for the critical eigenmode contain the same physics in the regime of interest. Indeed, we
are considering $\Delta\approx\de-\e\ll \de, \e$, such that at the end
of scaling $\de,\e>1$. Thus we can write $|\cL_{\al q}^R|^2\approx\de_\al^2$. 
Neglecting differences between $\de_b$ and $\de_t$ due to one-loop renormalizations, we
have $\langle\f_{\al q}\f_{\be q}^*\rangle \approx\langle\la_q\la_q^*\rangle/2$.
Hence, we will study the fluctuation dynamics in terms of the critical eigenmode, $\la$.

In equilibrium, Eq.(\ref{formalle}) in momentum-time space becomes
\beq
\label{leeq}
\frac{1}{v_F|{\b q}|}\partial_t\la({\b q},t)=
-(\De+|{\b q}|^2)\la({\b q},t)+\eta({\b q},t),
\eeq 
where we have dropped the $z=2$ term from the time-derivative because we are in the regime
of $|{\b q}|<\G/v_F$. 
In the quantum critical regime, the noise correlators become
\beq
\langle\eta({\b q},t)\eta({\b q'},t')\rangle\approx \frac{2T}{v_F|{\b q}|}\de(t-t')\de({\b q}+{\b q'}).
\eeq
We then find the noise-averaged correlation function for the fluctuations
\beq
\langle\la({\b q},t)\la(-{\b q},t')\rangle
\approx\frac{T}{\De}e^{-v_F|{\b q}|\De |t-t'|}.
\eeq

In the out of equilibrium case, the Langevin equation becomes
\bml
\frac{1}{v_F|{\b q}|}\partial_t\la({\b q},t)=
-\round{\De+|{\b q}|^2+\frac{i}{2}\b v_d\cdot{\b q}\tau}\la({\b q},t)+\eta({\b q},t).
\end{multline}
In the limit of large $T_{\rm eff}$, we find
\beq
\langle\eta({\b q},t)\eta({\b q'},t')\rangle\approx \frac{2T_{\rm eff}\tau}{\p}\de(t-t')\de({\b q}+{\b q'}).
\eeq
The noise-averaged correlation function then reads
\beq
\langle\la({\b q},t)\la(-{\b q},t')\rangle\approx
\frac{T_{\rm eff}\tau v_F|{\b q}|}{\p\De}e^{-v_F|{\b q}|(\De+\frac{i}{2}
{\b v_d}\cdot{\b q}\tau)|t-t'|}.\label{neq_corr}
\eeq

In equilibrium, the correlation function describes fluctuations which are conserved in 
nature with the decay rate vanishing in the long-wavelength limit as $\sim |{\b q}|$. Because
the effects of temperature are felt equally by both bosonic modes, the $z=2$
dynamics play no role in the long-wavelength theory of the order parameter dynamics.
The dynamics obeyed here is essentially identical to Model B in Ref. \onlinecite{handh}
but with the Landau damping parameter replaced by $v_F|\nabla|$.
The situation is different in the nonequilibrium case. Here, the delta-correlated noise
contribution only arises from $\Pi^K_b$ that reflects $z=2$ dynamics since the driving field 
(the source of noise) is
only coupled to the bottom electrons. The Langevin equation then displays a hybrid effect of both
$z=2$ and $z=3$ physics; while the white noise correlator reflects $z=2$ physics, the
damping is governed by the $z=3$ dynamical term. As a result, the corresponding
noise-averaged correlation function for the fluctuations yields a unique behaviour
where long-wavelength correlations vanish as $\sim |{\b q}|$. In addition, the advertised
drift of the flucutations can be explicitly seen in \req{neq_corr}.

\section{Conclusion}
\label{conclude}
In summary, we considered critical properties of a nonequilibrium bilayer system of itinerant electron
magnets. Starting from a microscopic fermionic model subject to an external drive, we derived
a coupled theory in terms of two bosonic fields which are related to the physical magnetization fluctuations
of the two layers. In the limit of no interlayer coupling, the fields obey different 
dynamics with different dynamical critical exponents ($z=2$ and $z=3$), leading to multi-scale quantum criticality
in the coupled system. We found that the applied current leads to both a drift of the magnetic fluctuations in the
coupled bilayer system and to decoherence. The latter phenomenon is more subtle in our system compared
with the analogous single layer scenario because the nonequilibrium drive is applied to only one of the layers. 
This causes it to play a role distinct from temperature in the thermal equilibrium case where both
layers are held at a common temperature. The differences are illustrated by comparing temperature
and nonequilibrium effects on the correlation length and the dynamics obeyed by the critical
fluctuations. A crucial feature of the work is that the two fields couple linearly. We found that 
the infrared properties of the system then are governed by the dynamics corresponding to the higher effective dimension. 
In light of Ref. \onlinecite{koln}, it would interesting to consider a coupled order parameter theory where 
the fields couple quadratically in the context of itinerant electron magnets. In this case, the effective 
theory possesses a discrete $Z_2\times Z_2$ symmetry corresponding to independent transformations 
$m_\al\rightarrow -m_\al$ for $\al=b,t$ ensuring that no linear coupling can be generated during the RG
transformation. If both fields become critical simultaneously, the low-energy properties are then expected 
to be governed by the field with the lower effective dimension.

\section*{Acknowledgments} 
The authors would like to thank Pawel Jakubczyk, Adrian Del Maestro, Aditi Mitra and Daniel Podolsky 
for helpful discussions, and Walter Metzner for a critical reading of the manuscript. 
This research was supported by NSERC of Canada, Canada Research Chair program, 
and Canadian Institute for Advanced Research (Y.~B.~K. and W.~W.-K.).

\begin{widetext}
\appendix
\section{Interlayer spin exchange}
\label{spinXc}
We now consider how the central insulator can mediate an effective spin exchange
interaction between the top and bottom layers. Imagine we have two 2D itinerant ferromagnets 
sandwiching a thin but 3D insulator with thickness $L$. For concreteness, we envisage the 
insulator as a quantum Ising paramagnet in the vicinity of a quantum phase transition to 
a long-ranged magnetically-ordered phase. The continuum quantum field theory for this model can 
be expressed in terms of a coarse-grained order 
parameter fluctuation field, $\phi$, subject to a confining potential. For simplicity, we truncate this 
potential at quadratic order with some invertible dynamic correlation matrix for the fluctuations, 
$\chi({\b x_i},{\b x_j},\tau)$, where ${\b x_i}=({\b r_i},z_i)$ labels the lattice sites in the 3D 
insulator. The QFT in the vicinity of the magnetic quantum critical point is then 
given by\cite{sachdev}
\beq
S_{\rm ins}=\int d\tau\sum_{ij}\phi({\b x}_i,\tau)\chi^{-1}({\b x}_i,{\b x}_j,\tau)\phi({\b x}_j,\tau).
\eeq
Since we are only interested in how this central insulator generates an effective spin exchange, we
will work in the equilibrium formalism for simplicity.

We now consider the interaction terms between the insulator and the two
itinerant magnetic layers. We assume that the plane of the bottom layer is $z=0$ and the top layer is at $z=L$.
The simplest plausible interaction would be a ferromagnetic
onsite exchange ($K>0$) between the layers and the insulator:
\begin{align}
S_{t-\rm ins}=-K\int d\tau\sum_{i}\hat{S}^z_t({\b r}_i,\tau)\phi({\b x}_i,\tau)\de_{z_i,L}, \\
S_{b-\rm ins}=-K\int d\tau\sum_{i}\hat{S}^z_b({\b r}_i,\tau)\phi({\b x}_i,\tau)\de_{z_i,0}.
\end{align}
Here, $\hat{S}^z_{t}$ and $\hat{S}^z_{b}$ are the Ising spin fluctuations in the
top and bottom layers, respectively. We integrate out the central fluctuations $\phi$
assuming that their in-plane correlations are short-ranged to obtain the following 
effective action
\beq
S_{\rm eff}=\int d\tau\sum_{\b r_i}\Big[-K^2\chi(L,L)S^z_t(\b r_i,\tau)
S^z_t(\b r_i,\tau) 
-K^2\chi(0,0)S^z_b(\b r_i,\tau)
S^z_b(\b r_i,\tau)
-J S^z_t(\b r_i,\tau)
S^z_b(\b r_i,\tau)\Big],
\eeq
where we have assumed that the inplane correlations of the insulator are short ranged and independent of 
$\tau$: $\chi(\b x_i,\b x_j,\tau)=\delta_{\b r_i,\b r_j}\chi(z_i,z_j)$. As advertised, 
an effective interlayer spin coupling $J=K^2[\chi(L,0)+\chi(0,L)]$ is generated.
We see that $-K^2\chi(L,L)$ and $-K^2\chi(0,0)$
lead to finite renormalizations of the intralayer exchange term.

\section{Polarization functions for the $\f_b$ sector}
\label{bottomPF}
In equilibrium, the retarded polarization function in the $\f_b$ sector is given by
\beq
\Pi^R_{bq}\approx -1+|\b q|^2- i\W\tau,
\eeq
where we are assuming $\W\tau<1$ and $q\tau v_F<1$. In the regime where $q\tau v_F\gg 1$, the dynamical
term becomes $\W/v_Fq$, i.e. conserved in nature. This is because on sufficiently short length scales the electrons
are unaffected by the presence of the substrate. The equilibrium contribution to the Keldysh polarization is given simply through the fluctuation-dissipation theorem,
\beq
\Pi^K_{bq}=-2i\abs{\W}\tau.
\eeq
To get the nonequilibrium corrections to the polarization functions we use 
the nonequilibrium Green's functions for the bottom electrons obtained in Sec.\ref{freefermionaction}.
From Eqs.(\ref{grab}),(\ref{gkbaditi}), the nonequilibrium contribution to the retarded polarization 
reads
\beq
\de\Pi^R_b(0,{\b q})\approx-i\frac{4\G^2}{\nu\p}
\int_{\b k}\frac{e{\b E}\cdot\b{v_k}\tau}{(\xi_{\mathbf{q+ k}}^2+\G^2)(\xi_{\b k}^2+\G^2)}.
\eeq
Expanding this result for small ${\b q}$ one finds
\beq
\de\Pi^R_b(0,\b q)\approx i\frac{1}{4m\G^2}e{\b E}\cdot{\b q}.
\label{drift}
\eeq
The nonequilibrium contribution to the Keldysh polarization is given by
\begin{multline}
\de\Pi^K_b(\W,{\b 0})
=\frac{-i}{\nu}\int_k\frac{-2i\G}{(\w+\W-\xi_{\b k})^2+\G^2}\frac{-2i\G}{(\w-\xi_{\b k})^2+\G^2}
\left\{-\sgn(\w+\W)\sgn(\w)e^{-\frac{\abs{\w}}{\abs{e{\b E}\cdot{\b v_F}\tau}}}\right. \\
\left. -\sgn(\w)\sgn(\w+\W)e^{-\frac{\abs{\w+\W}}{\abs{e{\b E}\cdot{\b v_F}\tau}}} 
+ [\sgn(\w)\sgn(\w+\W)+1]e^{-\frac{\abs{\w}}{\abs{e{\b E}\cdot{\b v_F}\tau}}} e^{-\frac{\abs{\w+\W}}{\abs{e{\b E}\cdot{\b v_F}\tau}}}\right\}.
\label{piknoneq}
\end{multline}
Here, we have linearized the fermion spectrum in the exponents.
For a weak field (i.e. $T_{\rm eff}\ll\tau^{-1}$)
the functions in $\{\dots\}$ in Eq.(\ref{piknoneq}) are strongly
peaked at $\w$ values where the numerators in the
exponents vanish. Making use of this we obtain
\beq
\de\Pi^K_b(\W,{\b 0})\approx \frac{4i\G^2}{\nu}\int_k\frac{1}{[\xi_{\b k}^2+\G^2]^2} \{ \dots \},
\label{piknoneq2}
\eeq
where $\{\dots\}$ is unchanged from Eq.(\ref{piknoneq}). This makes the $\w$-integral trivial, yielding the result 
\beq
\de\Pi^K_b(\W,{\b 0})=-i\frac{4\G^2}{\nu\p}\int\frac{kdkd\thi}{(2\p)^2}
\frac{T_{\rm eff}\abs{\cos\thi}
e^{-\frac{\abs{\W}}{T_{\rm eff}\abs{\cos\thi}}}}{[\xi_{\b k}^2+\G^2]^2},
\label{piknoneq3}
\eeq
where $\thi$ is the angle subtended by the Fermi velocity and the electric field. 
Performing the momentum integral gives
\beq
\de\Pi^K_b(\W,{\b 0})\approx-2i\tau\abs{\W} I\round{\frac{T_{\rm eff}}{\abs{\W}}}.
\eeq

\section{Evaluation of the correlation length}
\label{CLapp}
In this appendix, we present the derivation for the leading correction to the correlation length 
in the quantum critical regime. Here, the analysis is conducted in the $\phi_b$-$\phi_t$ basis.
The results are summarized in Secs. \ref{CLeq} and \ref{CLneq}. 

Recall that $\de_\al(b)$ in equilibrium is given by Eq.(\ref{desoleq}). The correction to the 
zero-temperature distance from criticality, $r$, is then given by
\beq
\De\de_\al=3u^\al_1\int_0^{\ln b^*}dxe^{-3x}[f_\al(T e^{3x},\tau e^{-x})-f_\al(0,\tau e^{-x})].
\eeq
If we define $\De f_\al:=f_\al(T e^{3x},\tau e^{-x})-f_\al(0,\tau e^{-x})$, we have 
\beq
\De f_\al=iK_2\int\frac{d\W}{2\p}\De D^K_\al(T,\tau),
\eeq
where $K_2=1/2\p$ and
\Beq
\De D^K_b(1,\W)&=&-2i\W\square{\coth\round{\frac{\W}{2T}}-\sgn(\W)}\frac{\tau[(\de+1)^2+\W^2]+\e^2}
{\abs{(\de+1-i\W\tau)(\de+1-i\W)-\e^2}^2};\\
\De D^K_t(1,\W)&=&-2i\W\square{\coth\round{\frac{\W}{2T}}-\sgn(\W)}\frac{(\de+1)^2+(\W\tau)^2+\tau\e^2}
{\abs{(\de+1-i\W\tau)(\de+1-i\W)-\e^2}^2}.
\Eeq
The $\W$-integral gets most of the contribution from $|\W|\le 2T$. So we can cutoff the integral, 
and approximate $\coth(\W/2T)\approx 2T/\W$. Then we obtain
\Beq
\label{dkk1}
\De D^K_b(1,\W)&\approx&-2i\round{2T-|\W|}\frac{\tau[(\de+1)^2+\W^2]+\e^2}
{\abs{(\de+1-i\W\tau)(\de+1-i\W)-\e^2}^2};\\
\label{dkk2}
\De D^K_t(1,\W)&\approx&-2i\round{2T-|\W|}\frac{(\de+1)^2+(\W\tau)^2+\tau\e^2}
{\abs{(\de+1-i\W\tau)(\de+1-i\W)-\e^2}^2}.
\Eeq
To scale up to $b^*$, where $T(b^*)=1$, we split the integral into two regimes: $1<b<b_1$, and $b_1<b<b^*$,
where $b_1=\de^{-1/2}$. Up to scale $b_1$, $\de,\e\ll 1$ and $T\ll 1$. So in this regime, we can use the
following expressions
\Beq
\label{dk1weakeq}
\De D^K_b&\approx&-2i\round{2T-|\W|}\frac{\tau}{1+(\W\tau)^2};\\
\label{dk2weakeq}
\De D^K_t&\approx&-2i\round{2T-|\W|}\frac{1}{1+\W^2}.
\Eeq
With this approximation, the $\W$-integral can be done trivially. Expanding the result for small $T$, we get
\Beq
\De f_b&=&\frac{4K_2}{\p}\tau T^2+\mathcal{O}(T^3),\\
\De f_t&=&\frac{4K_2}{\p}T^2+\mathcal{O}(T^3).
\Eeq
To evaluate the integral up to $b_1$, we first do a change of variables, 
$x\rightarrow T(b)=:y$, which makes the measure $dx=dy/3y$, and the lower and the upper limits 
$T $ and $T /\de^{3/2}$, respectively. Then the corrections are given by
\Beq
\De\de_\al^{(1)}&=&3u^\al_1\int_{T }^{T /\de^{3/2}}\frac{dy}{3y}\frac{T }{y}\De f_\al\round{y,\tau 
\round{\frac{T }{y}}^{1/3}};
\Eeq
We then obtain
\Beq
\label{dd1strongeq}
\De\de_b^{(1)}&=&\frac{3}{\p^2}u^b_1\tau T ^{4/3}\frac{T ^{2/3}}{\de},\\
\label{dd2strongeq}
\De\de_t^{(1)}&=&\frac{2}{\p^2}u^t_1T \frac{T^{2/3}}{\de}.
\Eeq

To scale between $b_1$ and $b^*$, we use Eqs.(\ref{dkk1}),(\ref{dkk2}) in the limit of $\de\sim\e\gg 1$.
Since $\De\ll 1$ throughout the flow, we note that $\de^2-\e^2\sim \de \De\le 1$ during the flow, and we drop this
term from the denominator. Then Eqs.(\ref{dkk1}),(\ref{dkk2}) can be approximated by
\Beq
\De D^K_b(1,\W)&\approx&-2i\round{2T-|\W|}\frac{\tau[\de^2+\W^2]+\e^2}{4\de^2+(\de\W(1+\tau))^2};\\
\De D^K_t(1,\W)&\approx&-2i\round{2T-|\W|}\frac{\de^2+(\W\tau)^2+\tau\e^2}{4\de^2+(\de\W(1+\tau))^2}.
\Eeq
We then find that
\Beq
\De f_b&\approx& \frac{\tau\de^2+\e^2}{4\p^2\de^2}T^2\approx \frac{1+\tau}{4\p^2}T^2,\\
\De f_t&\approx& \frac{\tau\e^2+\de^2}{4\p^2\de^2}T^2\approx \frac{1+\tau}{4\p^2}T^2,
\Eeq
where we have used the fact that $\de\approx \e$. The corrections from this region of the integral is then
given by
\beq
\De\de_\al^{(2)}=3u^\al_1\int_{T /\de^{3/2}}^{1}\frac{dy}{3y}\frac{T }{y}\De f_\al\round{y,\tau 
\round{\frac{T }{y}}^{1/3}}.
\eeq
We then obtain
\beq
\label{ddastrongeq}
\De\de_\al^{(2)}=\frac{u^\al_1T }{4\p^2}\square{1+\frac{3\tau }{2}T ^{1/3}}.
\eeq
Since $T^{2/3}\ll\de$ in this regime, Eqs.(\ref{dd1strongeq}),(\ref{dd2strongeq}) are sub-leading
to Eq.(\ref{ddastrongeq}). We therefore obtain to leading order
\beq
\De\de_\al\approx\frac{u^\al_1T }{4\p^2}\square{1+\frac{3\tau }{2}T^{1/3}}.
\eeq

In the nonequilibrium case, recall that $\de_\al(b)$ is given by Eq.(\ref{desolneq}). The correction to the 
zero-temperature distance from criticality, $r$, is then given by
\beq
\De\de_\al=3u^\al_1\int_0^{\ln b^*}dxe^{-3x}[f_\al(T _{\rm eff}e^{3x},\tau e^{-x})-f_\al(0,\tau e^{-x})].
\eeq
In this nonequilibrium case, we obtain
\Beq
\label{dk1neq}
\De D^K_b(1,\W)&=&-2i|\W| I\round{\frac{T_{\rm eff}}{|\W|}}\frac{\tau[(\de+1)^2+\W^2]}
{\abs{(\de+1-i\W\tau)(\de+1-i\W)-\e^2}^2};\\
\label{dk2neq}
\De D^K_t(1,\W)&=&-2i|\W| I\round{\frac{T_{\rm eff}}{|\W|}}\frac{\tau\e^2}
{\abs{(\de+1-i\W\tau)(\de+1-i\W)-\e^2}^2}.
\Eeq
In the above, the drift term was dropped since it does not reach cutoff scale at $b^*$.\cite{aditi2}

We first scale up to $b_1$. As in the equilibrium case, we have $\de\ll 1$ and $\e\ll 1$ in this region, and 
Eqs.(\ref{dk1neq}),(\ref{dk2neq}) can be approximated by
\Beq
\label{dk1weakneq}
\De D^K_b&\approx&\frac{-4i\tau T_{\rm eff}}{\p}\frac{1}{1+(\W\tau)^2};\\
\label{dk2weakneq}
\De D^K_t&\approx&\frac{-4i\tau T_{\rm eff}}{\p}\frac{\e^2}{[1+(\W\tau)^2](1+\W^2)}.
\Eeq
Once again, we have used the fact that $\De D^K_\al$ is appreciable only for $|\W|\le 2T_{\rm eff}$ and that
in this region,
\beq
I\round{\frac{T_{\rm eff}}{|\W|}}\approx\frac{T_{\rm eff}}{|\W|}\frac{2}{\p}.
\eeq
Performing the $\W$-integral up to $2T_{\rm eff}$, we then obtain to lowest order in $T_{\rm eff}$, 
\Beq
\De f_b&=&\frac{4K_2}{\p}\tau T_{\rm eff}^2+\mathcal{O}(T_{\rm eff}^3),\\
\De f_t&=&\frac{4K_2}{\p}\tau\e^2T_{\rm eff}^2+\mathcal{O}(T_{\rm eff}^3).
\Eeq
The corrections up to $b_1$ can then be computed, giving
\Beq
\label{dd1strongneq}
\De\de_b^{(1)}&=&\frac{3}{\p^2}u^b_1\tau T _{\rm eff}^{4/3}\frac{T _{\rm eff}^{2/3}}{\de},\\
\label{dd2strongneq}
\De\de_t^{(1)}&=&\frac{3}{2\p^2}u_1^t\tau \e^2T _{\rm eff}^{2/3}\frac{T _{\rm eff}^{4/3}}{\de^2}
\approx \frac{3}{2\p^2}u^t_1\tau T _{\rm eff}^2.
\Eeq
To scale in the region $b_1<b<b^*$, we use 
\Beq
\De D^K_b(1,\W)&\approx&\frac{-4i\tau T_{\rm eff}}{\p}\frac{\tau\de^2}{4\de^2+[\de\W(1+\tau)]^2};\\
\De D^K_t(1,\W)&\approx&\frac{-4i\tau T_{\rm eff}}{\p}\frac{\tau\e^2}{4\de^2+[\de\W(1+\tau)]^2}.
\Eeq
Performing the $\W$-integral and expanding for small $T_{\rm eff}$, we obtain
\beq
\De f_b\approx\De f_t\approx \frac{\tau T^2_{\rm eff}}{2\p^3},
\eeq
where we have again used the fact that $\de\approx\e$.
The corrections are then given by
\beq
\label{dd3strongneq}
\De\de_\al^{(2)}\approx\frac{3\tau u^\al_1}{4\p^3}T _{\rm eff}^{4/3}.
\eeq
As in the equilibrium case, the contributions from the first region of the integral 
(Eqs.(\ref{dd1strongneq}),(\ref{dd2strongneq})) are sub-leading to those from the second region 
(Eq.(\ref{dd3strongneq})). Therefore, to leading order, we obtain the following correction:
\beq
\De\de_\al\approx\frac{3\tau u^\al_1}{4\p^3}T _{\rm eff}^{4/3}.
\eeq

\section{Eigenmode analysis}
\label{eigenbasisanal}

\subsection{Determining the eigenmodes}\label{eigenmodes_determination}
In this appendix, we develop a critical theory in the basis of the eigenmodes. The renormalization group
analysis will be carried out in this basis, and the correlation length calculation will be reconsidered. 
We determine the eigenmodes of the system by diagonalizing the gaussian action Eq.(\ref{sgausseff}). 
What we find in the following is that the diagonalization procedure gives rise to two new bosonic modes
which possess different masses. While one becomes critical at $\de=\e$, the other mode remains gapped.
This allows one to integrate out the latter and arrive at an effective low-energy theory in terms of
one critical mode. This is in stark constrast with Ref. \onlinecite{koln} where both bosonic modes simultaneously
become critical.

To aid with the eventual block-diagonalization of the Keldysh action, we first diagonalize the
equilibrium action in the Matsubara formalism, then map the theory back to Keldysh space.
The Matsubara gaussian action reads,
\beq
\mathcal{S}^{(2)}_{\rm eq}=\int_q \rvec{\f_{bq}^*}{\f_{tq}^*}
\mat{\cL^{\rm eq}_{bq}}{-\e}{-\e}{\cL^{\rm eq}_{tq}}\cvec{\f_{bq}}{\f_{tq}},
\eeq
where $\cL^{\rm eq}_{bq}=\de+|{\b q}|^2+|\W|\tau$ and $\cL^{\rm eq}_{tq}=\de+|{\b q}|^2+|\W|/v_F|{\b q}|$
are the equilibrium inverse susceptibilities.
Expanding to lowest order with respect to the fluctuating parts
of the polarization functions, the eigenvalues of the action read $\frac{1}{2}(\cL^{\rm eq}_{bq}+\cL^{\rm eq}_{tq})\pm\e$.
The expansion is well-defined in the interlayer coupling regime considered here (i.e. $\de\gtrsim\e>0$).
If we now denote the eigenmodes by $\la_{1q}$ and $\la_{2q}$, the gaussian action then reduces
to
\beq
\label{sgaussdiag}
\mathcal{S}^{(2)}_{\rm eq}=\int\frac{d^2q}{(2\p)^2}\frac{d\W}{2\p} \rvec{\la_{1q}^*}{\la_{2q}^*}
\mat{(\cL^{\rm eq}_{bq}+\cL^{\rm eq}_{tq})/2-\e}{0}{0}
{(\cL^{\rm eq}_{bq}+\cL^{\rm eq}_{tq})/2+\e}
\cvec{\la_{1q}}{\la_{2q}}.
\eeq
We can now map this Matsubara action back into the Keldysh form by simply replacing both fields
by two-component Keldysh fields, i.e. $\la_{iq}\rightarrow\La_{iq}=\rvec{\la_{iq}^{\rm cl}}
{\la_{iq}^{\rm q}}^T$, and replacing each polarization function by its corresponding matrix. The Keldysh
gaussian action is then given by
\beq
i\mathcal{S}^{(2)}_{\rm eff}=-i\int\frac{d^2q}{(2\p)^2}\frac{d\W}{2\p} \rvec{\La_{1q}^\dagger}{\La_{2q}^\dagger}
\mat{(\hat\cL_{bq}+\hat\cL_{tq})/2-\e\hat\tau_x}{0}{0}{(\hat\cL_{bq}+\hat\cL_{tq})/2+\e\hat\tau_x}
\cvec{\La_{1q}}{\La_{2q}},
\eeq
where $\hat\cL_{\al q}=(\tilde U\nu)^{-1}\hat\tau_x+\hat\Pi_{\al q}$.
We now consider the quartic terms. Recall that in the $\f$ basis, the quartic interactions were
given by
\Beq
iS_{\rm eff}^{(4)}&=&\int d^3x\left\{ -i\square{u_1^b\round{\phi^{\rm cl}_b}^3\phi^{\rm q}_b
+u_3^b\phi^{\rm cl}_b\round{\phi^{\rm q}_b}^3}
+\square{u_2^b\round{\phi^{\rm cl}_b}^2\round{\phi^{\rm q}_b}^2
+u_4^b\round{\phi^{\rm q}_b}^4}\right\} + (b\leftrightarrow t),
\label{quartic2}
\Eeq
where we have written terms in the form in which $u_i^\al$ are all real. To transform the quartic terms
to the eigenbasis, we use the transformation matrix in the limit of $q\rightarrow 0$ since momentum and
frequency dependent parts generate terms that are (RG) irrelevant compared with the leading constant coefficients.
We then obtain
\beq
\cvec{\f_b^{\rm cl,q}}{\f_t^{\rm cl,q}}\approx\frac{1}{\sqrt{2}}\mat{1}{1}{1}{-1}\cvec{\la_1^{\rm cl,q}}
{\la_2^{\rm cl,q}}.
\eeq
Inserting this transformation into Eq.(\ref{quartic2}) gives the quartic terms in the
eigenbasis. 

The mass terms of the two eigenmodes in Eq.(\ref{sgaussdiag}) are $\de-\e$ and $\de+\e$ for $\La_{1}$ 
and $\La_{2}$,
respectively. This implies that while $\La_1$ becomes critical at $\de=\e$, $\La_2$ remains gapped.
We may therefore integrate out the gapped mode from the theory and obtain a single-component
effective action only in terms of the critical eigenfield. In the process of integrating out $\La_2$, we
only retain terms that are quadratic in the latter. Terms linear and quadratic in $\La_2$ emerge from
the quartic terms, however the former will
generate contributions beyond quartic order in the critical field. After these simplifications, the quartic
action becomes
\bml
iS_{\rm eff}^{(4)}=-i\int d^3x\square{\bar u_1\round{\la_1}^3\la^{\rm q}_1
+\bar u_3\la_1\round{\la^{\rm q}_1}^3}
+\int d^3x\square{\bar u_2\round{\la^{\rm cl}_1}^2\round{\la^{\rm q}_1}^2
+\bar u_4\round{\la^{\rm q}_1}^4}
-i\int_{q,q',k}\rvec{\la_{2q}^{{\rm cl}*}}{\la_{2q}^{{\rm q}*}}\mat{Q_{11}}{Q_{12}}{Q_{21}}{Q_{22}}
\cvec{\la_{2q'}^{{\rm cl}}}{\la_{2q'}^{{\rm q}}},
\end{multline}
with $Q_{11}=3\bar u_1\la_{1,k+q}^{\rm cl}\la_{1,-k-q'}^{\rm q}-i\bar u_2\la_{1,k+q}^{\rm q}\la_{1,-k-q'}^{\rm q}$.
Here, $\bar u_i=(u_i^b+u_i^t)/4$. 
$Q_{12}$, $Q_{21}$ and $Q_{22}$ are also terms quadratic in $\la_1$, however, we do not
write them explicitly here because they will not be necessary in the following discussion. 
Now performing the gaussian integral over $\La_2$, and expanding the resulting $\Tr\ln$ to linear-order 
in $\bar u_i$, we obtain
\beq
i\mathcal{S}^{(2)}_{\rm eff}=-i\int_q\La_{1q}^\dagger\square{(\hat\cL_{bq}+\hat\cL_{tq})/2-\e\hat\tau_x}\La_{1q}
-i\int_q\La_{1q}^\dagger\mat{0}{\eta_1}{\eta_1}{i\eta_2}\La_{1q},
\eeq
where
\beq
\eta_1=\frac{3\bar u_1}{2}\int_k iD^K_{2k},\qquad\qquad
\eta_2=\bar u_2\int_k iD^K_{2k},
\eeq
and
\beq
\mat{D^K_{2q}}{D^R_{2q}}{D^A_{2q}}{0}=-\square{(\hat\cL_{bq}+\hat\cL_{tq})/2+\e\hat\tau_x}^{-1}.
\eeq
Both $\eta_1$ and $\eta_2$ are real quantities, and we have used the fact that 
$D^R_{2{\b q}}(t,t)=D^A_{2{\b q}}(t,t)=0$.
We find that $\bar u_2 \rightarrow 0$ as $T\rightarrow 0$ so $\eta_2$ does not change the form of the Keldysh
term for $\La_1$. We will omit these terms. The $\bar u_1$ term gives a small renormalization to the mass 
of the critical mode. We absorb 
the $\eta_1$ into the new mass: $\De=\de-\e+\eta_1$. The full effective action for the critical mode is now given by
\beq
\label{effactioneq}
i\mathcal{S}_{\rm eff}=-i\int_q\La_{1q}^\dagger\mat{0}{\chi_A^{-1}}{\chi_R^{-1}}{\chi_K^{-1}}\La_{1q}
-i\int d^3x\square{\bar u_1\round{\la^{\rm cl}_1}^3\la^{\rm q}_1
+\bar u_3\la^{\rm cl}_1\round{\la^{\rm q}_1}^3}
+\int d^3x\square{\bar u_2\round{\la^{\rm cl}_1}^2\round{\la^{\rm q}_1}^2
+\bar u_4\round{\la^{\rm q}_1}^4}.
\eeq
In equilibrium, 
\Beq
&&\chi_R^{-1}=\De+|{\b q}|^2-\frac{i}{2}\round{\W\tau+\frac{\W}{v_F|{\b q}|}},\\
&&\chi_K^{-1}=-i\coth\round{\frac{\W}{2T}}\round{\W\tau+\frac{\W}{v_F|{\b q}|}},
\Eeq
and out of equilibrium,
\begin{align}
& \chi_R^{-1}=\De+|{\b q}|^2+\frac{i}{2}\b v_d\cdot{\b q}\tau-\frac{i}{2}\round{\W\tau+\frac{\W}{v_F|{\b q}|}},\\
& \chi_K^{-1}=-i\frac{|\W|}{v_F|{\b q}|}-i|\W|\tau\round{1+I\round{\frac{T_{\rm eff}}{|\W|}}}.
\end{align}
Note that $I(T_{\rm eff}/|\W|)\sim T_{\rm eff}/|\W|$ for $T_{\rm eff}>|\W|$, and $\coth(\W/2T)\sim T/\W$ for
$T>|\W|$. We then see that, similar to previous works\cite{aditi1,aditi2}, 
both $T$ and $T_{\rm eff}$ act as a mass
for the quantum fluctuations. At $T=0$ and $T_{\rm eff}=0$, $\chi_K^{-1}$ vanishes at low energies
as $\sim|\W|$. However, for $T\ne 0$ or $T_{\rm eff}\ne 0$, $\chi_K^{-1}\ne 0$ as $\W\rightarrow 0$, and
the theory effectively becomes classical. 
\subsection{RG analysis in eigenmode basis}\label{eigenmodes_RG}
We now perform a renormalization group analysis of the effective action, Eq.(\ref{effactioneq}).
Following a similar analysis as in the main text, we find the following set of RG equations to lowest (linear) order 
in the quartic coupling,
\Beq
\frac{dT_{\rm eff}(b)}{d\ln b}&=&3T_{\rm eff}(b),\\
\frac{dv_d(b)}{d\ln b}&=&2v_d(b),\\
\frac{d\tau(b)}{d\ln b}&=&-\tau(b),\\
\label{uflow}
\frac{d\bar u_i(b)}{d\ln b}&=&-\bar u_i(b)+\mathcal{O}([\bar u_j]^2),\\
\label{deltaflowe}
\frac{d\De(b)}{d\ln b}&=&2\De(b)+3\bar u_1(b) f(T_{\rm eff}(b),\tau(b)).
\Eeq
The function $f(T_{\rm eff}(b),\tau(b))$ is given by
\beq
f(T_{\rm eff}(b),\tau(b))=iK_2\int\frac{d\W}{2\p}\chi_K(1,\W),
\eeq
where $\chi_K(p)=-\chi_K^{-1}(p)/|\chi_R^{-1}(p)|^2$.
Solving these equations, we arrive at
\Beq
T_{\rm eff}(b)&=&T_{\rm eff}b^3,\\
v_d(b)&=&v_d b^2,\\
\tau(b)&=&\tau b^{-1},\\
\bar u_i(b)&=&\bar u_i b^{-1},\\
\label{desolneq2}
\De(b)&=&b^2[\De+3\bar u_1
\int_0^{\ln b}dxe^{-3x}f(T_{\rm eff}e^{3x},\tau e^{-x})].
\Eeq

An analogous renormalization group analysis in the equilibrium case gives
\Beq
T(b)&=&Tb^3,\\
\tau(b)&=&\tau b^{-1},\\
\bar u_i(b)&=&\bar u_i b^{-1},\\
\label{desoleq2}
\De(b)&=&b^2[\De+3 \bar u_1
\int_0^{\ln b}dxe^{-3x}f(Te^{3x},\tau e^{-x})].
\Eeq

We now recompute the correlation length in the eigenbasis.
The scaling stops when $\De(b_1)\sim 1$. In the quantum disordered regime, $T$ and $T_{\rm eff}$ is small enough
so that $T(b_1)\ll 1$ and $T_{\rm eff}(b_1)\ll 1$. In this case, the correlation length of the system can
be obtained by setting $T=0$ and $T_{\rm eff}=0$. One then obtains
\beq
\label{xiqr}
\xi^{-2}=\De+3\bar u_1\int_0^\infty dxe^{-3x}f(0,\tau e^{-x})=:r.
\eeq
The condition for the occurrence of the regime is obtained by requiring $T(b_1)<1$ where 
$b_1=r^{-1/2}$. Then the condition reads 
\beq
\label{qtoccrossover}
T<r^{3/2},\qquad T_{\rm eff}<r^{3/2}.
\eeq
The non-universal shift to the bare mass in Eq.(\ref{xiqr}) indicates that we are above the upper critical
dimension and that $\bar u_1$ is a dangerously irrelevant operator.

When the inequality in Eq.(\ref{qtoccrossover}) is violated, the critical fluctuations become classical
in nature and the system enters the quantum critical regime. Here we compute the leading corrections to the 
correlation length in this regime. We scale up to either $T(b^*)=1$ or $T_{\rm eff}(b^*)=1$. 
Recall that the solution for $\De(b)$ in equilibrium is given by Eq.(\ref{desoleq2}). The correction to the 
zero-temperature distance from criticality, $r$, is then given by
\beq
\de r=3\bar u_1\int_0^{\ln b^*}dxe^{-3x}[f(Te^{3x},\tau e^{-x})-f(0,\tau e^{-x})].
\eeq
If we define $\De f:=f(Te^{3x},\tau e^{-x})-f(0,\tau e^{-x})$, we have 
\beq
\De f=iK_2\int\frac{d\W}{2\p}\De \chi_K(T,\tau),
\eeq
and
\beq
\De \chi_K(1,\W)=-i\W\square{\coth\round{\frac{\W}{2T}}-\sgn(\W)}\frac{\tau+1}{(\De+1)^2+(\W\tau+\W)^2/4}.
\eeq
The $\W$-integral gets most of the contribution from $-2T\le\W\le 2T$. So we can cutoff the integral there, 
and use $\coth(\W/2T)\approx 2T/\W$. Then we obtain
\beq
\label{dk1}
\De \chi_K(1,\W)\approx -i\round{2T-|\W|}\frac{1+\tau}{1+(\W\tau+\W)^2/4},
\eeq
where we dropped $\De$ from the denominator since it remains small during the flow.
The $\W$-integral can now be done trivially. Expanding the result for small $T$, we get
\beq
\De f=\frac{1}{\p^2}(1+\tau) T^2+\mathcal{O}(T^3).
\eeq
To evaluate the integral up to $b^*$, we first do a change of variables, 
$x\rightarrow T(b)=:y$, which makes the measure $dx=dy/3y$. The upper limit of the integral is 1,
and we extend the lower limit down to 0. Then the correction is given by,
\beq
\de r=3\bar u_1\int_{0}^{1}\frac{dy}{3y}\frac{T}{y}\De f\round{y,\tau\round{\frac{T}{y}}^{1/3}}.
\eeq
We then obtain
\beq
\de r=\frac{\bar u_1}{\p^2}T\round{1+\frac{3\tau}{2}T^{1/3}}.
\eeq

We now move on to the nonequilibrium case.
Recall that the solution $\De(b)$ for the nonequilibrium case is given by Eq.(\ref{desolneq2}). The correction to the 
zero-temperature distance from criticality, $r$, is then given by
\beq
\de r=3\bar u_1\int_0^{\ln b^*}dxe^{-3x}[f(T_{\rm eff}e^{3x},\tau e^{-x})-f(0,\tau e^{-x})].
\eeq
We then obtain
\beq
\De \chi_K(1,\W)=-i|\W|\tau \frac{I\round{\frac{T_{\rm eff}}{|\W|}}}{1+(\W\tau+\W)^2/4},
\eeq
where we have dropped the drift and the mass terms from the denominator since they 
remain small during the flow\cite{aditi2}.
Once again, the $\W$-integral is appreciable only for $|\W|\le 2T_{\rm eff}$. In this region, we have
\beq
I\round{\frac{T_{\rm eff}}{|\W|}}\approx\frac{T_{\rm eff}}{|\W|}\frac{2}{\p}.
\eeq
Performing the $\W$-integral up to $2T_{\rm eff}$, we then obtain to lowest order in $T_{\rm eff}$, 
\beq
\De f=\frac{2}{\p^3}\tau T_{\rm eff}^2+\mathcal{O}(T_{\rm eff}^3).
\eeq
The mass correction is then given by
\beq
\de r=3\bar u_1\int_{0}^{1}\frac{dy}{3y}\frac{T_{\rm eff}}{y}\De f\round{y,\tau\round{\frac{T_{\rm eff}}{y}}^{1/3}}.
\eeq
Performing the  integral, we obtain
\beq
\de r=\frac{3\bar u_1}{\pi^3}\tau T_{\rm eff}^{4/3}.
\eeq

To summarize the results, the correlation length in equilibrium reads
\beq
\xi^{-2}_{\rm eq}=r+\frac{\bar u_1}{\p^2}T\round{1+\frac{3\tau}{2}T^{1/3}},
\eeq
while out of equilibrium, one obtains
\beq
\xi^{-2}_{\rm neq}=r+\frac{3\bar u_1\tau}{\p^3}T^{4/3}_{\rm eff}.
\eeq
This is in exact agreement with the results obtained in the main text 
(c.f. Eqs. (\ref{CLstrongeq}),(\ref{CLstrongneq})).
\vspace{5mm}
\end{widetext}



\begin{thebibliography}{99}
\bibitem{noise} A.G. Green, J.E. Moore, S.L. Sondhi, and A. Vishwanath, Phys. Rev. Lett.
\textbf{97}, 227003 (2006).
\bibitem{cond1} D. Dalidovich and P. Phillips, Phys. Rev. Lett. \textbf{93}, 027004 (2004).
\bibitem{cond2} A.G. Green and S.L. Sondhi, Phys. Rev. Lett. \textbf{95}, 267001 (2005).
\bibitem{curr} P.M. Hogan and A.G. Green, Phys. Rev. B \textbf{78}, 195104 (2008).
\bibitem{ks1} S. Kirchner and Q. Si, arXiv:0909.3925.
\bibitem{ks2} S. Kirchner and Q. Si, Phys. Rev. Lett. \textbf{103}, 206401 (2009).
\bibitem{aditi1} A. Mitra, S. Takei, Y.B. Kim, and A.J. Millis, Phys. Rev. Lett. \textbf{97}, 236808 (2006).
\bibitem{aditi2} A. Mitra and A.J. Millis, Phys. Rev. B \textbf{77}, 220404(R) (2008).
\bibitem{handh} P.C. Hohenberg and B.I. Halperin, Rev. Mod. Phys. \textbf{49}, 435 (1977). 
\bibitem{hertz} J.A. Hertz, Phys. Rev. B \textbf{14}, 1165 (1976).
\bibitem{millis} A.J. Millis, Phys. Rev. B \textbf{48}, 7183 (1993).
\bibitem{stoner} This is in constrast to the system studied in Ref. \onlinecite{aditi1} where a nonequilibrium renormalization 
to the Stoner criterion was obtained at the mean-field level. 
\bibitem{aditi3} A. Mitra, Phys. Rev. B \textbf{78}, 214512 (2008).
\bibitem{kamenev} A. Kamenev, condmat/0412296.
\bibitem{mahan} G.D. Mahan, \textsl{Many-Particle Physics}, Plenum Press, New York, (1990).
\bibitem{rammer} J. Rammer and H. Smith, Rev. Mod. Phys. \textbf{58}, 323 (1986).
\bibitem{roschrev} H. von L\"{o}hneysen, A. Rosch, M. Vojta, and P. W\"{o}lfle, Rev. Mod. Phys. 
\textbf{79}, 1015 ( 2007).
\bibitem{stewart} G.R. Stewart, Rev. Mod. Phys. \textbf{73}, 797 (2001).
\bibitem{BKV} D. Belitz, T.R. Kirkpatrick, and T. Vojta, Phys. Rev. B \textbf{55}, 9452 (1997).
\bibitem{chubukov} J. Rech, C. P\'{e}pin, and A.V. Chubukov, Phys. Rev. B \textbf{74}, 195126 (2006).
\bibitem{moriya} T. Moriya, \textsl{Spin Fluctuations in Itinerant Electron Magnets}, 
Springer-Verlag, Berlin (1985).
\bibitem{oganesyan} V. Oganesyan, S.A. Kivelson, and E. Fradkin, Phys. Rev. B \textbf{64},
195109 (2001).
\bibitem{koln} M. Zacharias, P. W\"{o}lfle, and M. Garst, Phys. Rev. B \textbf{80},
165116 (2009).
\bibitem{cenke} C. Xu, arXiv:0909.2647.
\bibitem{pepin} I. Paul, C. Pepin, and M.R. Norman, Phys. Rev. B \textbf{78}, 035109 (2008).
\bibitem{schmittmann1} C.C. Hill, R.K.P. Zia, and B. Schmittmann, Phys. Rev. Lett. \textbf{77},
514 (1996).
\bibitem{schmittmann2} B. Schmittmann, C.C. Hill, and R.K.P. Zia, Physica A \textbf{239},
382 (1997).
\bibitem{sachdev} S. Sachdev, \textsl{Quantum Phase Transitions}, Cambridge University
Press, Cambridge, England (1999).

\end{thebibliography}
\end{document}